\documentclass{article}
\usepackage[utf8]{inputenc}
\usepackage{graphicx, dcolumn, amsmath, amsfonts}
\usepackage{MjEpp, upgreek, tikz, dsfont}
\usepackage{geometry}
\usepackage{authblk}
\usepackage{subfiles}
\usepackage{hyperref}
\usepackage{url}
  
\title{Overview of the phase space formulation of quantum mechanics with application to quantum technologies}

\author[1]{Russell P Rundle\thanks{r.rundle@bristol.ac.uk}}
\affil[1]{School of Mathematics, Fry Building, University of Bristol, UK}

\author{Mark J Everitt}
\affil{Quantum Systems Engineering Research Group, Department of Physics, Loughborough University, Leicestershire LE11 3TU, United Kingdom}

\begin{document}
\maketitle

\begin{abstract}

The phase-space formulation of quantum mechanics has recently seen increased use in testing quantum technologies, including methods of tomography for state verification and device validation.
Here, an overview of quantum mechanics in phase space is presented.
The formulation to generate a generalized phase-space function for any arbitrary quantum system is shown, such as the Wigner and Weyl functions along with the associated $Q$ and $P$ functions.
Examples of how these different formulations have been used in quantum technologies are provided, with a focus on discrete quantum systems, qubits in particular.
Also provided are some results that, to the authors' knowledge, have not been published elsewhere. 
These results provide insight into the relation between different representations of phase space and how the phase-space representation is a powerful tool in understanding quantum information and quantum technologies.

\end{abstract}


\tableofcontents

\section{Introduction}

The phase-space formulation is just one of many approaches to consider quantum mechanics, where the well-known Schr\"odinger wave function and Heisenberg matrix formulations of quantum mechanics were the first to be devised in 1925.
During the same decade, Hermann Weyl realized that the structure of quantum mechanics closely followed the rules of group theory, from this realization he presented the, now called, Weyl transform, that takes a Hamiltonian in phase space and transforms it into a quantum mechanical operator.\Cite{Weyl1927}
It wasn't until 1932 that the inverse transform was presented by Wigner,\Cite{Wigner1932} taking a quantum wavefunction and representing it in phase space -- the Wigner function.

Subsequently, in the 1940s, both Gronewold\Cite{Groenewold1946} and Moyal\Cite{Moyal1949} then developed Wigner's phase-space function to fit around the earlier language Weyl used to represent quantum mechanics in phase space.
By developing the tools to transform any arbitrary operator to the phase-space representation by a taking the expectation value of some operator, known as the kernel, and describe the evolution of quantum mechanics in phase space. 
This created a complete theory of quantum mechanics as a statistical theory, as long as one only wished to consider systems of position and momentum.
The phase-space representation was missing an important aspect of quantum mechanics, the representation of finite quantum systems.
Using the language of Weyl, Groenewold, and Moyal, in 1956 Stratonovich then introduced the representation of discrete quantum systems into phase-space methods.\Cite{Stratonovich56}

This development went largely unnoticed, while the phase-space formulation of systems of position and momentum started to pick up popularity in the quantum optics community.
Where additions to the Wigner function were introduced, the Husimi Q-function\Cite{Husimi1940} and the Glauber-Sudarshan P-function\Cite{Glauber1963, Sudarshan1963}.
The development of the phase-space representation by Glauber in particular led to important contributions.
First the development of the displacement operator, and then the relation of this displacement operator to the generation of phase-space functions.\Cite{Glauber1969,Glauber1969-2}
The above mentioned kernel that maps an arbitrary operator to phase space was realized then discovered to be a displaced parity operator, the wording came from Royer\Cite{Royer1977}, where the $Q$ and $P$ functions have a similar construction, albeit with a different `parity'.
The displaced parity construction is an important discovery for much of this review and in the development of phase-space methods in experimental settings and quantum technologies in general.
However, the displaced parity formalism wasn't used for discrete quantum systems until more recently.\Cite{Amiet1,Amiet2}

In the late 1980s, interest in representing qubits, or even qudits, in phase space started to reemerge.
In 1987, Feynman\Cite{Feynman1987-FEYNP} and Wootters\Cite{WOOTTERS19871} both independently constructed a Wigner function for discrete systems that differs from the earlier construction from Stratonovich, however a relation between these constructions will be shown later. 
Feynman's work was a writing up of an earlier, back-of-the-envolope musings about negative probabilities, and only focuses on spin-1/2 systems.
Wootters on the other hand, was more interested in creating a Wigner function for any discrete system, where he created a Wigner function based on a discrete toroidal lattice that holds for dimension $d$ that is a prime power, due to the underlying Galois field structure.

Since this initial construction of a discrete phase space, there has been effort to extend this formulation to be useful more generally, see \Refs{Cohendet_1988, Leonhardt1995, PhysRevA.70.062101, Ruzzi_2005, Chaturvedi_2006, Gross2016} and \Refe{Ferrie2009} for an overview of this progress.
This also resulted in some key discoveries for finite systems in phase space, such as considering coupled qubits\Cite{Wootters2004,Paz2005}, qudits\Cite{Delfosse_2017}, and the link between Wigner function negativity and contextuality.\Cite{Raussendorf2017,Delfosse2015,Delfosse_2017,Schmid2018,Schmid2020,Schmid2021}

Shortly after Feynman and Wootters presented their findings, V\'arilly and Gracia-Bond\'ia came across Stratonovich's early work and brought it to public attention in 1989.\Cite{Varilly:1989gs}
Filling in some of the earlier gaps and also presenting the coherent-state construction\Cite{Arecchi1972, Perelomov, PerelomovB} to relate it more closely with Glauber's earlier work.
This was then followed by further development.\Cite{DowlingAgarwalSchleich, BrifMann1999,Amiet1,Amiet2,Klimov055303,Klimov2002,Cunha2001,Adam2020,koczor2019} 
These developments focussed on an \SU{2} structure of phase space, restricting larger Hilbert spaces to the symmetric subspace.
Alternatively, there has been development in describing larger systems in \SU{N}, although requiring many more degrees of freedom, can fully describe a quantum state.
The development of \SU{N} Wigner functions can be found in \Refs{TilmaKae1,TilmaEveritt,Rundle2018,Marchiolli_2019,Klimov-1008.2920}.

As quantum technologies advanced into the 21st century, direct measurement of phase space became increasingly more possible and useful.
Experiments then utilized on the displaced parity formulation to displace a state and make a direct parity measurement experimentally; for example the measurement optical states in QED experiments\Cite{Deleglise:2008gt}, and the rotation of qubit before taking a generalized parity measurement in \Refe{Rundle2016}.
More examples of such experiments are given within this review.

In this review, we begin by presenting the formulation of quasiporbability distribution functions for systems of position and momentum, where we will then generalize this formulation for use in any arbitrary system in \Sec{Formulation}.
This will provide an understanding of the general framework needed to understand how to map a given operator to its representation in phase space, and how such a mapping is bijective and informationally complete.
Following this in \Sec{FiniteSystems}, we will use this generalized formulation to introduce the phase-space formulation of various finite-dimensional quantum systems.
This will include an in-depth discussion on different representations of qubits in phase space and how different representations are related; we will also present multiple ways one can represent multi-qubit states in phase space dependent on different group structures.
We will then describe how these methods can be used in practice for use in quantum technologies in \Sec{QTinPS}; where we will translate some important figures of merit for quantum computation into the language of phase space, as well discussing the importance of negative values in the Wigner function.
This will be followed by examples of experiments performed on qubits where the phase space distributions were directly measured.

\section{Formulation of phase space}\label{Formulation}

In this section we will lay the groundwork for how one can generate a phase-space distribution for any arbitrary operator, where we will provide a general framework for any group structure in \Sec{GenFramework}.
We will begin with the original formulation in position-momentum space, within the Heisenberg-Weyl group (HW) where we will provide the general formula to create any quasiprobability distribution function and any characteristic function.
This will be followed by looking into the dynamics of such states and how representing quantum mechanics in phase space allows one to see to what extent one can treat dynamics classically.
This section will end with a general framework and a summery of the results in \Refe{Rundle2018}.

\subsection{Systems of position and momentum} \label{HWWignerSection}

Wigner's original formulation was generated in terms of a wavefunction, $\psi(q)$, and is in essence the Fourier transform of a correlation function, where for a function defined in terms of position and momentum
\begin{equation}
	\HWWig_{\psi}(q,p) = \int_{-\infty}^\infty \psi^*\!\left(q+\frac{z}{2} \right)\psi\!\left(q-\frac{z}{2} \right) \ue{\ui pz} \ud z,
\end{equation}
where we will assume that $\hbar = 1$.
We note that in the original construction, the Wigner function was presented to describe the position-momentum phase space for many particles, here we will restrict the discussion to one concomitant pair of degrees of freedom where we will discuss the composite case later in this section.

This formulation of the Wigner function was then soon generalized to apply to any arbitrary operator, $\OpA$, as an expectation value of a kernel, $\HWPi(\alpha)$, such that\Cite{Groenewold1946,Moyal1949}
\begin{equation}\label{WignerFunctionHW}
	\HWWig_A(\HWvar) = \Trace{\OpA \; \HWPi(\HWvar)},
\end{equation}
where we now define the Wigner function in terms of the complex variable $\alpha = \left(q +\ui p\right)/\sqrt{2}$ to fit inline with the quantum optics literature.

The kernel in \Eq{WignerFunctionHW} is fully constructed as a displaced parity operator
\begin{equation}\label{HWKernel}
	\HWPi(\HWvar) = \HWD(\HWvar) \,\HWPi \, \HWD^\dagger(\HWvar),
\end{equation}
which consists of the modified parity operator 
\begin{equation}\label{HWParity}
	\HWPi = 2\ue{\ui\pi\Opad\Opa},
\end{equation}
which is two-times the standard parity operator, where $\Opa$ and $\Opad$ are the annihilation and creation operators respectively.
The second component of \Eq{HWKernel} is the displacement operator
\begin{equation}\label{HWDisplacement}
	\HWD(\HWvar) = \exp\left(\HWvar\Opad  - \HWvar^*\Opa \right) = \ue{-\frac{1}{2} |\HWvar|^2}\ue{\HWvar\Opad}\ue{-\HWvar^*\Opa} = \ue{\frac{1}{2} |\HWvar|^2}\ue{-\HWvar^*\Opa}\ue{\HWvar\Opad}.
\end{equation}
The displacement operator displaces any arbitrary state around phase space and is central to the definition of a coherent state,\Cite{Glauber1963,Perelomov,Zhang1990} which is generated by displacing the vacuum state, $\ket{0}$, such that
\begin{equation}
\HWD(\HWvar) \ket{0} = \ket{\HWvar}	,
\end{equation}
where $\ket{\alpha}$ is an arbitrary coherent state.
The coherent states form an overcomplete basis, where the resolution of identity requires integration over the full phase space
\begin{equation}
	\frac{1}{\pi}\int_{-\infty}^\infty \OP{\alpha}{\alpha} \,\ud^2\alpha = \Bid,
\end{equation}
and the overlap between any two coherent states is non-zero, where
\begin{equation}
	\IP{\alpha_1}{\alpha_2} = \exp\left[ \frac{1}{2}\left(|\alpha_1|^2 +|\alpha_2|^2 - 2\alpha_1^*\alpha_1\right) \right].
\end{equation}

Note that there is a factor of 2 in \Eq{HWParity}, and subsequently in \Eq{HWKernel}, this is chosen for two reasons.
First, it aligns with the formulation of the Wigner function that was introduced by Glauber when writing it in a displaced parity formalism.
But more importantly, it lines up more nicely with the definition of the Wigner function for discrete systems.
Later, we will introduce a condition that requires the trace of the kernel to be one, which requires \Eq{HWParity} to have the factor of 2 out the front to hold.

The Wigner function is just one of a whole class of potential quasiprobability distribution functions for quantum systems, what makes it a preferable choice is that the kernel to transform to the Wigner function is the same as the kernel used to transform back to the operator, which is not always the case, such that
\begin{equation}\label{HWReverseWignerTransform}
	\OpA = \frac{1}{\pi} \int_{-\infty}^\infty \HWWig_A(\HWvar)\, \HWPi(\HWvar) \ud \HWvar,
\end{equation}
showing that the Wigner function is an informationally complete transformation of any arbitrary operator.
Another important advantage is that the marginals of the Wigner function are the probability distribution of the wavefunction.
That is for the Wigner function of a state $\psi(q)$
\begin{equation}\label{marginals}
	|\psi(q)|^2 = \int_{-\infty}^{\infty} W(q,p) \, \ud p,
\end{equation}
integrating over all the values of $p$ to produce the probability for $\psi(q)$, likewise for $\psi(p)$ when integrating over $q$.

To represent quantum mechanics in phase space, some properties one would usually desire from a probability distribution must be sacrificed -- hence the prefix `quasi' in quasiprobability distribution functions.
For the Wigner function this sacrifice is on positive-definiteness, a more in-depth conversation on the negativity of the Wigner function will be given in \Sec{WigNeg}.
If however one doesn't want to sacrifice a non-negative probability distribution function, the alternative is to consider the Husimi $Q$ function\Cite{Husimi1940}, where instead the sacrifice comes in the form of losing the correct marginals.

The $Q$ function is generated in terms of coherent states, where
\begin{equation}
	Q_A(\HWvar) = \left \langle \HWvar \left | \OpA \right| \HWvar \right \rangle = \Trace{\OpA \ket{\HWvar}\!\bra{\HWvar}} = \Trace{\OpA \, \left(\HWD(\HWvar)\ket{0}\!\bra{0}\HWD^\dagger(\HWvar)\right)},
\end{equation}
where the last form of the $Q$ function can be thought of as an expectation value of a general coherent state, where instead of the kernel being a displaced parity operator it is a displacement of the projection of the vacuum state.

Another drawback to the $Q$ function for some is that transforming back to the operator will require a different kernel, this is the kernel that transforms an arbitrary operator to the Glauber-Sudarshan $P$ function\Cite{Glauber1969,Sudarshan1963}.
Likewise, the kernel to transform back from the $P$ function to the operator is the kernel for the $Q$ function -- this is the usual definition of the $P$ function
\begin{equation}
	\OpA  = \int_{-\infty}^\infty P_A(\HWvar) \, \ket{\HWvar}\!\bra{\HWvar} \, \ud \HWvar  = \int_{-\infty}^\infty P_A(\HWvar) \, \left(\HWD(\HWvar)\ket{0}\!\bra{0}\HWD^\dagger(\HWvar) \right)\, \ud \HWvar,
\end{equation}
where $P_A(\HWvar)$ is the $P$ function for an arbitrary operator, $\OpA$.
The $P$ function seems to sacrifice more, as it doesn't have the correct marginals and it is not non-negative, in fact coherent states in the P functions are singular, which may be an attractive analogy when considering coherent states as so-called `classical stats', although we must note that coherent states are still fundamentally quantum.
Despite the apparent drawbacks, there have been a set of key results by analysis of the $P$ function, see for example \Refe{PhysRevA.92.033837} -- this will also be discussed further in \Sec{QTinPS}

Apart from the Wigner, $Q$ and $P$ functions, there are further choices of quasi-probability distribution function for quantum systems.
In order to completely describe the full range of possible phase-space functions, including the appropriate kernels to transform both to and from the given function, it is first necessary to introduce the quantum mechanical characteristic functions.
The characteristic function is a useful tool in mathematical analysis, and is the Fourier transform of the probability distribution function.
In the same way, the characteristic functions in quantum mechanics are Fourier transforms of the quasi-probability distribution functions.
The Fourier transform of the Wigner function is also known as the Weyl function and is defined
\begin{equation}\label{HWWeylFunction}
	\HWWeyl_A(\HWWeylvar) = \Trace{\OpA \, \HWD(\HWWeylvar)},
\end{equation}
such that
\begin{equation}
	\HWWeyl_A(\HWWeylvar) = \frac{1}{\pi} \int_{-\infty}^\infty \HWWig_A(\HWvar)\,\exp\left(\HWvar\HWWeylvar^* - \HWvar^*\HWWeylvar \right)\, \ud \HWvar, \hspace{0.5cm} \text{and,} \hspace{0.5cm} \HWWig_A(\HWvar) = \frac{1}{\pi} \int_{-\infty}^\infty \HWWeyl_A(\HWWeylvar)\,\exp\left(\HWWeylvar\HWvar^* - \HWWeylvar^*\HWvar \right)\, \ud \HWWeylvar,
\end{equation}
relate the Wigner and Weyl functions.

The superscript in brackets $(0)$ in \Eq{HWWeylFunction} is a specific case of the characteristic function, $\HWChar{s}_A(\HWvar)$, for $-1\leq s \leq 1$; the Weyl function is just the special case when $s=0$.
In general a quantum characteristic function is defined
\begin{equation}\label{HWGenChar}
	\HWChar{s}_A(\HWWeylvar) = \Trace{\OpA \, \HWD_s(\HWWeylvar)}, \hspace{0.5cm} \text{where,} \hspace{0.5cm} \HWD_{s}(\HWWeylvar) = \HWD(\HWWeylvar)\ue{\frac{1}{2}s\left|\HWWeylvar\right|^2}.
\end{equation}
Note that when $s=0$, $\HWD_0(\HWWeylvar)$ is the standard displacement operator from \Eq{HWDisplacement}.
When $s=0$ we say that the displacement operator is symmetically, or Weyl, ordered.
Alternatively, when $s=-1$ or $s=1$ the corresponding displacement operators, and therefore characteristic functions, are anti-normal and normal respectively; this can be realized by showing the operators explicitly
\begin{equation}
	\HWD_{-1}(\HWWeylvar) =  \ue{-\HWvar^*\Opa}\ue{\HWvar\Opad} \hspace{0.5cm} \text{and} \hspace{0.5cm} \HWD_{1}(\HWWeylvar) = \ue{\HWvar\Opad}\ue{-\HWvar^*\Opa},
\end{equation}
demonstrating the opposite order of the arguments in the exponents.
Note that, like the quasi-probability distribution functions, the characteristic functions are informationally complete, such that
\begin{equation}
	\OpA = \int_{-\infty}^\infty \HWChar{s}_A(\HWWeylvar) \, \HWD_{-s}^\dagger(\HWWeylvar) \, \ud\HWWeylvar,
\end{equation}
where the reverse transform kernel is the Hermitian conjugate of the generalized displacement operator with minus the $s$ value of the function.

The $Q$ and $P$ functions are then the Fourier transforms of the anti-normally ordered and normally ordered characteristic functions respectively, where
\begin{equation}
	Q(\HWvar) = \frac{1}{\pi} \int_{-\infty}^\infty \HWChar{-1}_A(\HWWeylvar)\,\exp\left(\HWWeylvar\HWvar^* - \HWWeylvar^*\HWvar \right)\, \ud \HWWeylvar, \hspace{0.5cm} \text{and,} \hspace{0.5cm} P(\HWvar) = \frac{1}{\pi} \int_{-\infty}^\infty \HWChar{1}_A(\HWWeylvar)\,\exp\left(\HWWeylvar\HWvar^* - \HWWeylvar^*\HWvar \right)\, \ud \HWWeylvar.
\end{equation}
More generally we can say that they are the normally and anti-normally quasi-probability distribution, where we can define a general $s$-values quasi-probability distribution function
\begin{equation}\label{HWGeneralPDF}
	F^{(s)}_A(\HWvar) = \frac{1}{\pi} \int_{-\infty}^\infty \HWChar{s}_A(\HWWeylvar)\,\exp\left(\HWWeylvar\HWvar^* - \HWWeylvar^*\HWvar \right)\, \ud \HWWeylvar,
\end{equation}
which hold for any value of $s$, generating an array of quasi-probability distribution functions, however only $s=-1,0,1$ will be considered here.

Further, by using the definition of the characteristic function in \Eq{HWGenChar} and substituting it into \Eq{HWGeneralPDF}, we can see that by bringing the integral inside the trace operation we can define the generalized kernel for any quasi-probability distribution function as the Fourier transform of the displacement operator
\begin{equation}
	\HWPi^{(s)}(\HWvar) = \frac{1}{\pi} \int_{-\infty}^\infty \HWD^{(s)}(\HWWeylvar)\exp\left(\HWWeylvar\HWvar^* - \HWWeylvar^*\HWvar \right)\, \ud \HWWeylvar,
\end{equation}
Note this can alternatively be expressed as weighted integrals of the displacement operator, see \Refe{koczor2019} for example.
Where a general function in HW phase space is defined as
\begin{equation}
	F^s_A(\HWvar) = \Trace{\OpA\, \HWPi^{(s)}(\HWvar)}.
\end{equation}
For the reverse transform, to take a phase-space distribution to an operator, one can then perform
\begin{equation}\label{HWReverseTransform}
	\OpA =  \int_{-\infty}^\infty F^{(s)}_A(\HWvar) \,\HWPi^{(-s)}(\HWvar) \,\ud \HWvar,
\end{equation}
providing the transforms between an arbitrary density operator and an arbitrary $s$-valued quasiprobability distribution function.
To complete the picture, all that is needed is a discussion of the dynamics on quantum systems in phase space.

\subsection{Evolution in phase space}

\newcommand{\PB}[2]{\left\{#1,#2\right\}}
\DeclarePairedDelimiter{\MbBraces}{\{\!\{}{\}\!\}}
\newcommand{\MB}[2]{\MbBraces{#1,#2}}
\newcommand{\CHamiltonian}{\mathscr{H}}
\newcommand{\Tends}{\rightarrow}

Dynamics of quantum systems are particularly appealing in the phase space formulation.
To understand why this is the case we return to classical Hamiltonian physics and understand the phase space approach in terms of the Wigner function from this perspective.
The classical dynamics of any distribution $A$ for a given Hamiltonian $\CHamiltonian$ is given by
\begin{equation} \label{eq:HamiltonianDynamics}
  \TD{A}{t} = \PB{A}{\CHamiltonian} + \PD{A}{t},
\end{equation}\enlargethispage{12pt}
where the Poisson bracket
\begin{equation}
  \PB{f}{g}= \sum_i \PD{f}{q_i}\PD{g}{p_i}-\PD{f}{p_i}\PD{g}{q_i},
\end{equation}
defines the way that a quantities flow in phase space; as the Poisson bracket is an example of a Lie bracket, the fact that this represents a specific kind of infinitesimal transform for classical dynamics is natural. 
The Poisson bracket thus specifies the way in which the state of the system transforms incrementally in time for a given Hamiltonian as a flow in phase space. For example, in a time increment $\ud t$ the coordinates
\begin{equation}( q_i, p_i)\rightarrow (q_i+\PB{q_i}{\CHamiltonian}  \ud t, p_i+ \PB{p_i}{\CHamiltonian}  \ud t),\end{equation}
and we can see how the Hamiltonian defines a ``flow'' in phase space (the above picture illustrates this in one-dimension). If we change the definition of the Poisson bracket this would result in different physics and the phase space formulation of quantum mechanics follows exactly this line of reasoning. In this way we will see that in redefining the Poisson bracket Quantum mechanics can be seen as a continuous deformation of this underlying "flow" algebra.

Before proceeding we first note that if we set $A$ equal to the probability density function $\rho(t)$ then we observe two important corollaries. The first is that conservation of probability requires $\TD{\rho}{t}=0$ (so it is a  constant, but not necessarily an integral, of the motion) and the second is that $\PB{\rho}{\CHamiltonian}$ can be thought of as the divergence of the probability current. This is expressed as in the Liouville ``conservation of probability'' equation (which has obvious parallels with the von-Neumann equation for density operators): 
\begin{equation}
    \PD{\rho}{t}=\PB{\CHamiltonian}{\rho}.
\end{equation}
When Dirac motivated the Heisenberg picture of quantum mechanics he did so by arguing that equations of the above form and  the Lie algebraic structure of the Poisson bracket be both retained, if quantum dynamics is to be defined in terms of some infinitesimal transform this is a sensible requirement. 
Dirac's argument then was to introduce other mathematical structures in which to achieve this end -- operators in a vector space. 

The alternative we take here is the Deformation Quantization approach that leaves all quantities in exactly the same form as in classical  Hamiltonian physics but to introduce quantum phenomena by deforming the phase space directly. 
In this view, function multiplication is replaced by a `star' product, such that
$$\lim_{\hbar \Tends 0} f \star g = f g$$
and that we introduce an alternative to the Poisson bracket - termed the Moyal bracket 
$$
 \MB{f}{g} = \frac{1}{{\ui}\hbar}\left[ f \star g - g \star f\right],
$$
that satisfies the property
\begin{equation}
  \PB{f}{g}= \lim_{\hbar \Tends 0} \MB{f}{g}.
\end{equation}
This last expression guarantees that any such formulation of `quantum mechanics' will smoothly reproduce classical dynamics in the limit $\hbar \Tends 0$ and for this reason can be viewed as a deformation of classical physics. 
For this reason we briefly reintroduce $\hbar$ in this section for comparison with classical mechanics, we will then return to $\hbar=1$ in the next section.
Understanding the Wigner function as the exact analog of the probability density function -- and as a quantum constant of the motion -- means that conservation of probability requires $\TD{W}{t}=0$, we also note that $\MB{W}{\CHamiltonian}$ can be thought of as the divergence of this generalized probability current.
The Louiville equation is then replaced by the more general equation:
\begin{equation}\label{eq:quantumLouiville}
    \PD{W}{t}=\MB{\CHamiltonian}{W}.
\end{equation}
We see that stationary states are then those for which $\MB{\CHamiltonian}{W}=0$ which lead to left and right $\star$-genvalue equations that are the phase space analog of the time-independent Schr\"odinger equation (whose solutions will be integrals of the motion as the phase-space analogue of energy eigenstates). 
Although this is not a foundations paper it is  worth noting: (i) the quantisation of phase space resolves issues around operator ordering in convectional quantum mechanics (see Groenewalds theorem for example\Cite{Groenewold1946}); (ii) 
 that this approach may be of particular value when looking to unify quantum mechanics and gravity as time is treated on the same footing as all other coordinates and the geometric view of introducing curvature through the metric tensor is perfectly natural in a phase space formulation.

The star-product is usually written, for canonical position and momentum, in its differential form as
\begin{equation}\label{theStarProduct}
  f \star g = f \sum_{i=1}^N \exp \left[\ui \hbar 
  \left(\overset{\leftarrow}{\PD{}{q_i}}\overset{\rightarrow}{\PD{}{p_i}}- \overset{\leftarrow}{\PD{}{q_i}}\overset{\rightarrow}{\PD{}{p_i}}\right)\right] g
\end{equation}
where the arrows indicate the direction in which the derivative is to act. Which can be shown to be:
\begin{align}
f (q,p) \star g (q,p)
&=
f \left(q+\frac{\ui \hbar}{2} \overrightarrow{\partial_p},p-\frac{\ui \hbar}{2} \overrightarrow{\partial_q} \right)  g (q,p)
\label{SP:EASY1}
\\
&=f (q,p)g \left(q-\frac{\ui \hbar}{2} \overleftarrow{\partial_p},p+\frac{\ui \hbar}{2} \overleftarrow{\partial_q} \right)
\label{SP:EASY2}
\\
&=f \left(q+\frac{\ui \hbar}{2} \overrightarrow{\partial_p},p\right)g \left(q-\frac{\ui \hbar}{2} \overleftarrow{\partial_p},p \right)
\label{SP:EASY3}
\\
&=f \left(q,p-\frac{\ui \hbar}{2} \overrightarrow{\partial_q}\right)g \left(q,p+\frac{\ui \hbar}{2} \overleftarrow{\partial_q} \right) \label{SP:EASY4}
\end{align}
where $(q+\frac{\ui \hbar}{2} \overrightarrow{\partial_p})$ and $(p-\frac{\ui \hbar}{2} \overrightarrow{\partial_q})$ are known as the Bopp operators.\Cite{Bopp}

Importantly for our discussion  the star product can also be written in a convolution-integral form:
\begin{equation}
f(q,p) \star g(q,p) = \frac{1}{\pi^2 \hbar^2} \, \int dq' dp' dq'' dp'' \, f(q+q',p+p') \, g(q+q'',p+p'') \, \exp{
\left(\frac{2\ui}{\hbar}
\left(q'p''-q''p'\right)\right)
}.
\end{equation}
which has the advantage that it both highlights that quantum effects are non-local and, from a computational perspective, numerical integration can be less prone to error than differentiation (although more costly).

We now can use a specific example to a result by Brif and Mann\Cite{BrifMann1999} to connect this form to a more general expression
\begin{equation}
f(q,p) \star g(q,p) =  \int dq' dp' dq'' dp'' \, f(q',p') \, g(q'',p'') \, \Trace{\HWPi(q,p)\HWPi(q',p')\HWPi(q'',p'')}.
\end{equation}
where $\HWPi(q,p)$ is the usual displaced parity operator. 


To summarise the quantum dynamics of any system or composite of systems is given by 
\begin{align}\label{HWEvolution}
    \PD{W}{t}&=\MB{\CHamiltonian}{W} 
    \nonumber \\
    \MB{W}{\CHamiltonian} &= \frac{1}{{\ui}\hbar}\left[ W \star \CHamiltonian - \CHamiltonian \star W\right]
    \nonumber \\
    W(\Omega) \star 	\CHamiltonian(\Omega) &=	 \int_{\Omega''}\int_{\Omega'} W(\Omega')\CHamiltonian(\Omega'') \, K(\Omega',\Omega'';\Omega) \, \ud \Omega' \ud\Omega'' 
\end{align}
where we see this as an equivalent view to classical mechanics with a deformation of function multiplication that in the limit $\hbar \Tends 0$ will, if the limit exits, reduce these dynamics to classical physics.
This can be viewed as moving from a non-local to local theory as $W(\Omega) \star 	\CHamiltonian(\Omega) \Tends W(\Omega)  	\CHamiltonian(\Omega)$ as $\hbar \Tends 0$.

We note that owing to the marginals of the Wigner function producing a positive distribution, one can formulate quantum dynamics in terms of the marginal distribution of the shifted and squeezed quadrature components. 
Thus producing an alternative formulation of the quantum evolution.\Cite{Mancini1996}

As a final note the classical dynamics of particles should even be recoverable in this limit as $\hbar \Tends 0$ the Wigner function of, e.g., coherent states tends to Dirac delta functions centred at one point in phase space $\rho(q_i,p_i)=\prod_i\delta(q_i-Q_i,p_i-P_i)$ and the Liouville equation for point particles will recover that classical point trajectories in an analogous way to solutions of the Kontsevich equation used in plasma physics.\Cite{Kontsevich2003}
As Moyal brackets tend to Poisson brackets as $\hbar \Tends 0$ it is natural to see that the Kontsevich equation is recoverable in this limit.

\subsection{Quantum mechanics as a statistical theory}\label{GenFramework}

We will now generalize the discussion of phase space for use for any group structure.
This will include us generalizing important equations from \Sec{HWWignerSection} and providing the generalization for the evolution in \Eq{HWEvolution}.

To generalize the formulation of phase-space functions, a suitable candidate for the given group structure needs to be chosen to replace the Heisenberg-Weyl displacement operator.\Cite{Rundle2018}
Such a generalized displacement operator, $\GenD(\Genvar)$ then displaces the corresponding choice of vacuum state for the given system, $\Genvac$, creating the generalized coherent state in that system\Cite{PerelomovB,Zhang1990}
\begin{equation}\label{GenCoherentState}
	\ket{\Genvar} = \GenD(\Genvar)\Genvac.
\end{equation}
For examples of such operators and the generation of generalized coherent states and the generation of generalized displacement operators for different systems, see for example \Refs{PerelomovB,Tilma1,Tilma2,Tilma3,TilmaKae1,MByrd1,MByrdp1}.
Note that when generating quasiprobability distribution functions, any suitable generalized displacement operator can produce an informationally complete function, however when considering just the Weyl characteristic function much more care needs to be taken in considering the correct operator,\Cite{Yan1985,Rundle2018} since it is crucial to determine a generalized displacement operator that generates an informationally complete function, where the inverse transform to the operator exists, such that
\begin{equation}\label{GenWeylFunction}
	\GenChar{s}_A(\GenWeylvar) = \Trace{\OpA \, \GenD_s(\GenWeylvar)} \hspace{0.5cm} \text{where,} \hspace{0.5cm} \OpA = \int_{\GenWeylvar}  \GenChar{s}_A(\GenWeylvar) \GenD_{-s}^\dagger(\GenWeylvar)\,\ud\GenWeylvar,
\end{equation}
where $\ud \GenWeylvar$ is the volume-normalized differential element or Haar measure for the given system.

Once a generalized displacement operator has been chosen, the generation of the $Q$ and $P$ functions simply requires the generalized coherent state from \Eq{GenCoherentState}, where
\begin{equation}\label{QandPfunctions}
	Q_A(\Genvar) = \left\langle \Genvar \left| \OpA \right| \Genvar \right\rangle \hspace{0.5cm} \text{and,} \hspace{0.5cm} \OpA = \int_{\Genvar} P_A(\Genvar) \, \ket{\Genvar}\!\bra{\Genvar} \, \ud\Genvar,
\end{equation}
however the construction of the Wigner function is somewhat more involved.

To generate the Wigner function we now need to construct the corresponding generalized parity operator, $\GenPi$, to create the kernel $\GenPi(\Genvar)$, yielding the Wigner function
\begin{equation}\label{GeneralWignerKernel}
	\GenWig_A(\Genvar) = \Trace{\OpA \; \GenPi(\Genvar)}.
\end{equation}
In \Refe{Stratonovich56}, Stratonovich set out five criteria that such a kernel needs to fulfil  to generate a Wigner function, known as the Stratonovich-Weyl correspondence.
These are, taken and adapted from \Refe{TilmaEveritt}:

{\renewcommand{\labelenumi}{\footnotesize{\upshape{S-W.~\theenumi}}}
\begin{minipage}{0.95\textwidth}
\begin{quote}
\begin{enumerate}
\item\label{SW1} Linearity:
	The mappings $\GenWig_{A}(\Genvar)=\Trace{\OpA \, \GenPi(\Genvar)}$ and $\OpA = \int_{\Genvar} \GenWig_{A}(\Genvar) \GenPi(\Genvar) \ud \Genvar$ exist and are informationally complete. 
	This means that $\OpA$ can be fully reconstructed from $\GenWig_{A}(\Genvar)$ and vice versa. 
\item\label{SW2} Reality:
	$\GenWig_{A}(\Genvar)$ is always real valued when $\OpA$ is Hermitian, therefore $\GenPi(\Genvar)$ must be Hermitian. 
	From the structure of the kernel in \Eq{GeneralWignerKernel}, this also means that the generalized parity operator, $\GenPi$, must also be Hermitian.
\item\label{SW3}  Standardization:
	$\GenWig_{\OpA}(\Genvar)$ is `standardized' so that the definite integral over all space $\int_{\Genvar} W_{A}(\Genvar) \ud \Genvar = \Trace{\OpA}$  exists and $\int_{\Genvar} \GenPi(\Genvar) \ud \Genvar =\Bid$.
\item\label{SW4} Traciality:
	The definite integral $\int_{\Genvar} W_{A}(\Genvar)W_{B}(\Genvar) \ud \Genvar = \Trace{\OpA \OpB} $ exists.  
\item\label{SW5} Covariance:
	Mathematically, any Wigner function generated by `rotated' operators $\GenPi(\Genvar^{\prime})$ (by some unitary transformation $V$) must be equivalent to `rotated' Wigner functions generated from the original operator ($\GenPi(\Genvar^{\prime}) \equiv V \GenPi(\Genvar) V^{\dagger}$) - \textit{i.\ e.\ }if $\OpA$ is invariant under global unitary operations then so is $\GenWig_{A}(\Genvar)$.
\end{enumerate}
\end{quote}
\end{minipage}
}

Note that from S-W.~\ref{SW1}, the transform back to the operator
\begin{equation}\label{ReverseTransformGeneral}
	\OpA = \int_{\Genvar} \GenWig_{A}(\Genvar) \GenPi(\Genvar) \; \ud \Genvar
\end{equation}
is integrated over the full volume $\Genvar$, and $\ud\Genvar$ is the volume normalized differential element, see Appendices in \Refe{TilmaKae1} or \Refe{Rundle2018} for more details.
Also note that, given the kernel for a system, the generalised parity can be recovered by taking
\begin{equation}\label{Pi00}
	\GenPi = \GenPi(\boldsymbol{0}),
\end{equation}
this is a helpful result in future calculations.\Cite{Rundle2018, Amiet1, Amiet2}
We can therefore say in general that any quasi-probability distribution function is defined
\begin{equation}\label{GenPhaseSpaceFunction}
	F^{(s)}_A(\Genvar) = \Trace{\OpA\, \GenPi^{(s)}(\Genvar)},
\end{equation}
where 
\begin{equation}
	\GenPi^{(s)}(\Genvar) = \GenD(\Genvar) \GenPi^{(s)} \GenD^\dagger(\Genvar)
\end{equation}
is the generalized kernel for any $s$-valued phase-space function.

As is demonstrated from \Eq{ReverseTransformGeneral}, the Wigner function is an informationally complete representation of the state, \textit{i.~e.}~it contains all the same information as the density operator as one can transform between the two representations without loss of information.
The same can be said of the $Q$ and $P$ functions, where the operator can be returned from the $Q$ function by integrating over the kernel for the $P$ function, likewise the operator can be returned by integrating the $P$ function over the $Q$ function kernel, explicitly
\begin{equation}\label{GenReverseTransform}
	\OpA = \int_{\Genvar} F^{(s)}_A(\Genvar) \GenPi^{(-s)}(\Genvar)  \,\ud \Genvar,
\end{equation}
which is the generalization of \Eq{HWReverseTransform}.

It is now natural to ask whether there is a simple comparison between the different phase-space functions without first going to the density operator.
Indeed this can be done by way of a generalized Fourier transform.\Cite{BrifMann1999,Varilly:1989gs,Rundle2018,Koczor2020}
We can simply bypass the calculation of the density operator by substituting \Eq{GenReverseTransform} for one value of $s$ into \Eq{GenPhaseSpaceFunction} for a different $s$ value, yielding
\begin{equation}\label{GenFourierTransform}
\begin{split}
	F^{(s_2)}_A(\Genvar') &= \Trace{ \int_{\Genvar} F^{(s_1)}_A(\Genvar) \GenPi^{(-s_1)}(\Genvar)  \,\ud \Genvar \, \GenPi^{(s_2)}(\Genvar')}\\
	&= \int_{\Genvar} F^{(s_1)}_A(\Genvar) \Trace{ \GenPi^{(-s_1)}(\Genvar) \, \GenPi^{(s_2)}(\Genvar')}  \,\ud \Genvar \\
	&= \int_{\Genvar} F^{(s_1)}_A(\Genvar) \,\mathcal{F}(\Genvar',s_2;\Genvar,s_1)  \,\ud \Genvar
\end{split}
\end{equation}
where
\begin{equation}
	\mathcal{F}(\Genvar',s_2;\Genvar,s_1) = \Trace{ \GenPi^{(-s_1)}(\Genvar) \, \GenPi^{(s_2)}(\Genvar')},
\end{equation}
is the generalized Fourier operator.
This can further be generalized to transform between any phase-space function, including the Weyl function, by substituting in \Eq{GenWeylFunction} instead.

We can further extend the idea behind this Fourier kernel to perform a convolution between any two phase-space functions to create a third function, such that
\begin{equation}
	F^{(s)}_{AB}(\Genvar) = \int_{\Genvar'}\int_{\Genvar''} F^{(s_1)}_{A}(\Genvar')F^{(s_2)}_{B}(\Genvar'') \mathcal{K}(\Genvar',\Genvar'';\Genvar) \,\ud\Genvar'\ud\Genvar'',
\end{equation}
where 
\begin{equation}\label{GenConvultion}
	\mathcal{K}(\Genvar',\Genvar'';\Genvar) = \Trace{\GenPi^{(s)}(\Genvar)\,\GenPi^{(-s_1)}(\Genvar')\GenPi^{(-s_2)}(\Genvar'')},
\end{equation}
is the convolution kernel.
This then provides a generalization of \Eq{theStarProduct} allowing us to define the evolution of any general system by\Cite{Rundle2018} 
\begin{equation}
	\PD{W_\rho}{t} = -\ui \left[W_{\rho H}(\Omega) - W_{H \rho}(\Omega) \right],
\end{equation}
for specific examples of the application of this equation on different quantum systems, see \Refe{WOOTTERS19871, Varilly:1989gs, BrifMann1999, Klimov055303, Klimov2002, Mancini1996, Koczor2020, KoczorPhD, Koczor2019Time, Koczor2019Continuous}.

Stratonovich showed in \Refe{Stratonovich56} that the kernels can be combined for the individual systems to create a kernel for a composite system.
The procedure to do so is straight-forward, it simply requires one to take the Kronecker product of the individual systems
\begin{equation}\label{CompositeKernel}
	\GenPi(\boldsymbol{\Genvar}) = \bigotimes_{i=1}^n \GenPi_i(\Genvar_i),
\end{equation}
for $n$ subsystems, where $\boldsymbol{\Genvar} = \{ \Genvar_1, \Genvar_2, ..., \Genvar_n \}$ and $\Genvar_i$ are the degrees of freedom for the individual systems.
Similarly $\GenPi_i(\Genvar_i)$ is the kernel for each individual system.
The full composite Wigner function is then generated by using \Eq{CompositeKernel} in \Eq{GeneralWignerKernel}.

Transforming back to the operator from the composite Wigner function requires a simple modification of \Eq{ReverseTransformGeneral}, where the new volume normalized differential operator is 
\begin{equation}
	\ud \boldsymbol{\Genvar} = \prod_{i=1}^n \ud\Genvar_i,
\end{equation}
more details on this construction can be found in \Refe{Rundle2018}.
We now have all the ingredients necessary to generate a quasi-distribution functions for any system, or any composition of individual systems, to represent quantum technologies in phase space.



\section{Phase-space formulation of finite quantum systems}\label{FiniteSystems}

We will now turn our attention to the generation of phase space functions for finite quantum systems.
It will be shown how the general framework from \Sec{GenFramework} can be applied to specific finite-dimensional quantum systems to represent them in phase space.
Where consideration of such structures is important in considering the use of phase-space methods in quantum technologies.
As such, we will show the important case of representing qubits in phase space, providing a link between the two most well-known formulations of qubits in phase space.
This will be followed by looking at how more general qudits and collections of qubits can be represented in phase-space in different situations.
Given the phase-space representation of all these systems, along with the continuous-variable construction for optical quantum systems, we will then be able to consider many approaches to quantum technologies in phase space.

\subsection{Qubits in phase space}

The most important system to introduce now for use in quantum technologies is the construction of phase-space functions for qubits.
The qubit Wigner function has two main formulations, there is the original construction by Stratonovich that considers the qubit over the Bloch sphere, with continuous degrees of freedom.
The second formulation was generalized by Wootters, and considers the qubit states on a discrete toroidal lattice, with discrete degrees of freedom.

Both formulations will be considered here, as each has had important results in understanding the properties of quantum mechanics in a quantum technology setting.
These two formulations are also specific examples of a general formulation of qudits in phase space.
It is also interesting to note that the discrete-lattice qubit Wigner function from Wootters can be considered a sub-quasiprobability distribution of the continuous-spherical Wigner function.
This relation will be shown later, but this relation does not hold for general qudit states.

We will also consider the generation of the $Q$ and $P$ functions for qubits, which are already well-known in the Stratonovich formulation, however to our knowledge does not exist in the discrete case.

\subsubsection{The Stratonovich kernel}

\begin{figure}
  	\includegraphics[width =\linewidth]{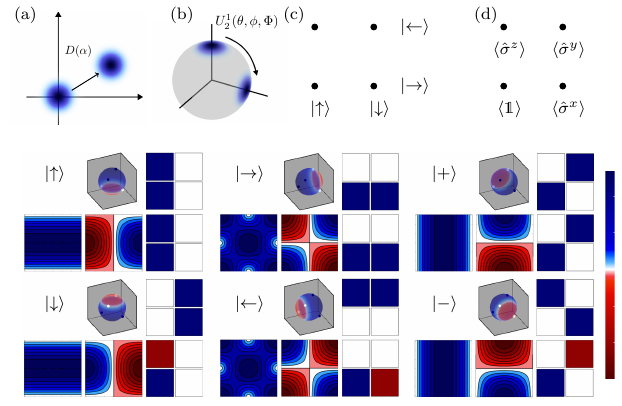}
	\caption{\label{QubitExamples} Qubits in phase space.
	(a) shows the action of the displacement operator on the vacuum state in phase space for the Heisenberg-Weyl group.
	When considering qubits, the geometry of the phase space is different, where two geometries can be chosen to represent qubits.
	(b) shows a spherical phase space, similar to the Bloch sphere. 
	Instead of the displacement operator in (a), a rotation operator is needed to rotate a state around the sphere.
	Alternatively, (c) and (d) show how qubit states can be represented on discrete toroidal lattices. 
	(c) shows how the entries for the discrete Wigner function and (d) is the discrete phase space for the discrete Weyl function. 
	Examples of the qubit Wigner function for the eigenstates of the Pauli matrices are shown below, where each state is marked with the ket representation of the state. 
	where $\ket{\uparrow}$ and $\ket{\downarrow}$ are the eigenstates of $\Sz$, $\ket{\rightarrow}$ and $\ket{\leftarrow}$ are the eigenstates of $\Sx$ and $\ket{+}$ and $\ket{-}$ are the eigenstates of $\Sy$ with eigenvalues $\pm 1$ respectively.
	For each of the states, the top row shows the Wigner function in Stratonovich formalism on a sphere and then the Wootters formalism on a $4 \times 4$ grid.
	Four points are highlighted on the Wigner function generated with the Stratonovich kernel that correspond to the values of the discrete Wootters formulation.
	Below we show three plots of the Weyl function, where the we have the real and imaginary values of the continuous Weyl function respectively.
	This is followed by the discrete Weyl function.}
\end{figure}

Following the steps from \Sec{Formulation}, a suitable generalized displacement operator and generalized parity operator is necessary for the construction of a Wigner function for qubits.
We will start by considering an appropriate generalized displacement operator.
Since we are now concerned with distributions on the sphere, the generalized displacement operator is a rotation operator.
An example of the change in geometry and in displacement can be seen in \textbf{\FigSub{QubitExamples}{b}}.
A complicating factor in choosing an operator is that rotation around a sphere isn't a unique procedure, there are many ways in which one can perform such an operation.
Out of the options, we highlight two choices that are heavily used in the appropriate literature and give equivalent results upon rotation of the generalized parity operator.
These are 
\begin{equation}\label{SU2QubitRotations}
	\SUND{2}{1}(\phi,\theta,\Phi) = \exp\left(\ui\phi\Sz/2\right)\exp\left(\ui\theta\Sy/2\right)\exp\left(\ui\Phi\Sz/2\right),\hspace{0.5cm} \text{and } \hspace{0.5cm} \OpR(\xi) = \exp\left(\xi\OpSp - \xi^*\OpSm \right),
\end{equation}
which are the standard Euler angles, where we introduce the general notation $\SUND{N}{M}$ that in this case is set to \SU{N}, $M = 2j$, for the quantum number $j$. 
The second operator $\OpR(\xi)$, where $\xi = \theta\exp(-\ui\phi)/2$ and $\OpSpm = (\Sx \pm \ui \Sy)/2  $, is a generalization of the Heisenberg-Weyl displacement for \SU{2}, and can be found in much of the spin-coherent state literature -- for example see \Refs{PerelomovB, Arecchi1972} -- and can be related to the Euler angles by\Cite{Rundle2018}
\begin{equation}\label{qubitCoherentRot}
	\OpR(\xi) = \SUND{2}{1}(\phi,\theta,-\phi).
\end{equation}

Although it is much more tempting to have a preference towards this rotation operator, due to the similarity to the Heisenberg-Weyl case, this rotation operator generates a Weyl function that is not informationally complete.
For instance 
\begin{equation}\label{informationalIncompleteness}
	\bra{\uparrow}\OpR(\xi)\ket{\uparrow} = \bra{\downarrow}\OpR(\xi)\ket{\downarrow}.
\end{equation}
We therefore need to construct an informationally complete Weyl function by including all three Euler angles,\Cite{Rundle2018}
\begin{equation}\label{SWWeyl}
	\chi_A(\tilde{\phi},\tilde{\theta},\tilde{\Phi}) = \Trace{\OpA\,\SUND{2}{1}(\tilde{\phi},\tilde{\theta},\tilde{\Phi})},
\end{equation}
and so we prefer to keep this $\Phi$ in the definition of the rotation operator.
Further, keeping all three Euler angles allows us to directly relate to operations on a quantum computer, where it is usual to include an arbitrary rotation in terms of Euler angles -- or at least some operation with three degrees of freedom.
This is also the reason we prefer to use $\SUND{2}{1}(\phi,\theta,\Phi)$ over $\OpR(\xi)$, despite the latter having the satisfying connection to $\HWD(\alpha)$.

The reason for the requirement of all three angles can be thought of by considering the Weyl function as giving the expectation value of the rotation operator used.
When considering the rotation operators $\OpR(\xi)$, the rotation is on \CP{1} rather than the full \SU{2}.
This means we have rotations restricted to a 2-sphere.
In turn, the rotation operator cannot produce the state $\Sz$, which is necessary for discerning its two eigenstates $\ket{\uparrow}$ and $\ket{\downarrow}$.
This is the reason for the result in \Eq{informationalIncompleteness}.

It is interesting to note that when our goal is to use the Weyl function to consider the expectation value of different operators, we can instead consider the Weyl function on a torus, where we set $\Phi=0$.
It turns out that this restriction is informationally complete and the last degree of freedom is not strictly necessary in the qubit case, as long as the range is extended to $-\pi<\tilde{\phi},\tilde{\theta}\leq \pi$.
However, we will generally choose keep the $\Phi$ degree of freedom to allow the rotation operator to also define the coherent state representation on \CP{1}, and more importantly to keep this formulation informationally complete for larger Hilbert spaces.

We now need to use our choice of rotation operator to rotate the corresponding generalized parity operator.
For the Wigner function, this is explicitly
\begin{equation}\label{SU2Parity}
	\SUNPi{2}{1} =  \frac{1}{2} \left( \Bid + \sqrt{3} \Sz \right),
\end{equation}
where $\Sz$ is the standard Pauli $z$ operator.
The appearance of the $\sqrt{3}$ in \Eq{SU2Parity} may seem somewhat strange, as is generally assumed that the generalized parity should simply be $\Sz$, without the identity or the coefficients.
Choosing $\Sz$ as a parity operator can be found in some works, and has proven useful in verification protocols, such as in~\Refe{Song2019}, this choice, however, leads to a function that is not informationally complete, \textit{i.~e.}~\Eq{ReverseTransformGeneral} does not hold.
Note also that S-W.~\ref{SW3} implies that $\text{Tr}[\GenPi] = 1$, which does not hold for $\Sz$.
To generate a generalized parity that is informationally complete, it is therefore necessary to follow the Stratonovich-Weyl correspondence S-W.~\ref{SW1}~-~\ref{SW5}.
For more detail of how this can be done, see \Refe{Stratonovich56} or in the Appendix \Refe{RundlePhD}.
This yields the Stratonovich qubit Wigner function kernel
\begin{equation}
	\SUNPi{2}{1}(\theta,\phi) =  \SUND{2}{1}(\phi,\theta,\Phi) \SUNPi{2}{1} \SUND{2}{1}(\phi,\theta,\Phi)^\dagger,
\end{equation}
where the $\Phi$ degree of freedom gets cancelled out due to the generalized parity being a diagonal operator, this is the reason behind the equivalence of the rotation operators in \Eq{SU2QubitRotations} when rotating the generalized parity operator.

We can then simply return the operator by applying this case to \Eq{GenReverseTransform}
\begin{equation}
	\OpA  = \frac{1}{2\pi} \int_0^{2\pi}\int_0^{\pi}W_A(\theta,\phi)  \SUNPi{2}{1}(\theta,\phi) \sin\theta \, \ud\theta\,\ud\phi.
\end{equation}
If one, on the other hand, insisted on using just the $\Sz$ operator, creating a phase-space function $W^Z_\rho(\Omega) = \Trace{\DO\,\SUND{2}{1}(\Omega)\Sz\SUND{2}{1}(\Omega)^\dagger  }$, to transform back to the density operator, the identity needs to be reinserted
\begin{equation}\label{sZParity}
	\DO = \frac{1}{2}\Bid + \frac{1}{2\pi} \int_0^{2\pi}\int_0^{\pi}W^Z_\rho(\theta,\phi)  \Sz(\theta,\phi) \sin\theta \, \ud\theta\,\ud\phi,
\end{equation}
however, we therefore have to assume that the function is of a trace-1 operator, and this doesn't hold for an arbitrary operator.
This result is clearly shown by considering either \Refe{Stratonovich56} or \Refe{RundlePhD} and setting the coefficient of the identity operator to 0.
Although this can be useful for examining certain properties of a quantum state, it is important to bear in mind that it doesn't satisfy the Stratonovich-Weyl correspondence.
Further, using the \Sz operator as a parity creates a function similar to the Weyl function -- that is, it is the expectation value of some arbitrary operator in phase space.
The only problem is the identity operator is not represented in this distribution like the \Sz operator in not represented in the case of the \CP{1} rotation operator.

Examples of the single-qubit Wigner function for the eigenstates of the Pauli operators can be seen in \FigSub{QubitExamples}{d}, where $\ket{\uparrow}$ and $\ket{\downarrow}$ are the eigenstates of $\Sz$ with eigenvalues $\pm1$ respectively.
Likewise, $\ket{\rightarrow}$ and $\ket{\leftarrow}$ are the eigenstates of $\Sx$ with eigenvalues $\pm1$ and $\ket{+}$ and $\ket{-}$ are the eigenstates of $\Sy$ with eigenvalues $\pm1$.
To the right of each function that is generated by the Stratonovich kernel is the same state generated by the Wootters kernel, that will be shown in the following section.
More comparison between the two is therefore to follow.
It is immediately clear that when generated with the Stratonovich kernel on a spherical geometry, the Wigner function is similar to the more familiar Bloch sphere, where the highest value of the probability distribution is where the Bloch vector would point.
It is also worth noting that, unlike the Heisenberg-Weyl functions and further the discrete Wigner functions, every spin coherent state has a negative region.
This is a result of the spherical geometry and the role of quantum correlations within phase space, this will be discussed in more detail in \Sec{WigNeg}.

Alternatively, one may be interested in constructing the $Q$ or $P$ function for the qubit, where the generation of the suitable kernel is just as simple as in the qubit case.
Like the Wigner function generalized parity operator, these too are diagonal operators.
The two can be calculated explicitly as
\begin{equation}\label{QubitQPParity}
	\SUNPi{2}{1(-1)} =\frac{1}{2} \left( \Bid + \Sz \right) = \begin{pmatrix} 1 & 0 \\ 0 & 0 \end{pmatrix} ,\hspace{0.5cm} \text{and } \hspace{0.5cm}  \SUNPi{2}{1(1)} = \frac{1}{2} \left( \Bid + 3 \Sz \right) = \begin{pmatrix} 2 & 0 \\ 0 & -1 \end{pmatrix} 
\end{equation}
for the $Q$ and $P$ function respectively, where the $\pm1$ in the bracket signifies the value of $s$ for the generalized phase-space function.
These both result in kernels where the $\Phi$ degree of freedom also cancels out when taking a similarity transform with the rotation operator.
The cancellation of the $\Phi$ term can therefore mean that the last Euler angle is not necessary for the calculation of quasi-probability distribution functions, and one will occasionally see the kernel generated ignoring this term, for good reason to exclude redundant information.
This also lines up nicely with the toroidal definition of the Weyl function, where we only care about the first two degrees of freedom in the rotation operator.
As we will soon see, a toroidal geometry is a natural alternative to the Bloch sphere in considering qubit states.

\subsubsection{The Wootters kernel}

The alternative formulation to consider qubits in phase space can be seen in the framework set out by Wootters.\Cite{WOOTTERS19871} 
In the same year, Feynman also gave an identical construction of a Wigner function for qubits.\Cite{Feynman1987-FEYNP}
Feynman's construction treated only the qubit case, whereas in the work by Wootters the formalism was extended to any prime-powered dimensional phase space.
More importantly, in \Refe{WOOTTERS19871} Wootters presented the discrete Wigner function in the Moyal formalism -- the expectation value of a kernel, which Wootters intuitively names the `phase-point operator'.
Despite this, we will give the formulation of the discrete Wigner function presented by Feynman, as we feel it provides a different insight into the problem, and then subsequently relate it to Wootters's phase-point operator.
Following this, we will show how this formulation can be considered a sub-quasiprobability distribution of the Stratonovich formulation.

The idea behind Feynman's formulation was to write down probability distributions
\begin{eqnarray}\label{FeynmanProbabilities}
\begin{split}
	p_{++} &=& \frac{1}{2}\EX{\Bid + \Sx + \Sy + \Sz  }, \hspace{2cm} p_{+-} = \frac{1}{2}\EX{\Bid - \Sx - \Sy + \Sz  }, \\
	p_{-+} &=& \frac{1}{2}\EX{\Bid + \Sx - \Sy - \Sz  }, \hspace{2cm} p_{--} = \frac{1}{2}\EX{\Bid - \Sx + \Sy - \Sz  },
\end{split}
\end{eqnarray}
which are the joint probabilities, $p_{\pm\pm}$, of finding the qubit having the spin aligned with the $\pm z$ and $\pm x$ axes respectively.
The choices of operator and expectation values in \Eq{FeynmanProbabilities} were presented by Feynman without any derivation or motivation; although, as we will see later Wootters uses the same basis, this choice is not unique, in fact there is no unique choice of a probability distribution of this form, see \Refs{Scully1983, Cohen1986,Scully1994} for examples of generalizations of these results.

Following the results of Wootters, we can take the expressions in \Eq{FeynmanProbabilities} and turn them into a phase-point operator.
We first introduce two degrees of freedom to span the discrete phase space, these are $ z,x \in \{0,1\} $ which cover all four phase points.
The phase-point operator is then defined
\begin{equation}
	\DWPi_2(z,x) = \frac{1}{2} \left( \Bid +(-1)^z\,\Sz + (-1)^x\,\Sx + (-1)^{z+x}\,\Sy \right),
\end{equation}
resulting in the discrete Wigner function for any arbitrary operator
\begin{equation}
	\DWig_A(z,x) = \Trace{\OpA \; \DWPi_2(z,x)},
\end{equation}
where the subscript is dimension $d=2$ for $\DWPi_d(z,x)$.
We can then relate the results in \Eq{FeynmanProbabilities} by noting that $\DWig(0,0) = p_{++}$, $\DWig(0,1) = p_{+-}$, $\DWig(1,0) = p_{-+}$, and $\DWig(1,1) = p_{--}$.

Like the Stratonovich kernel, we can define a generalized displacement operator to both generate a Weyl function and further to be used to construct the Wootters kernel in terms of a generalized displaced parity operator.
The displacement operator for discrete systems such as this is
\begin{equation}\label{DiscreteDisplacement}
	\DWD_2(z,x) = \exp\left(\frac{1}{2}\ui\pi xz\right)(\Sx  )^x(\Sz)^z,
\end{equation}
which, like the \SU{2} Wigner function, can be used to generate a discrete Weyl characteristic function
\begin{equation}\label{DiscreteQubitWeyl}
	\DWeyl(\tilde{z},\tilde{x}) = \Trace{\OpA \, \DWD_2 (\tilde{z}, \tilde{x})}.
\end{equation}
The four values of the discrete displacement operator are simply the Pauli and identity operators.
As such, there is a direct relationship between the Euler-angle parameterization and toroidal lattice approach.

Note first that by setting $\Phi=0$ in the \SU{2} case, the rotation operator defines translations around a torus.
This then allows us to describe the values of the toroidal lattice as points in the continuous torus generates by the Euler angles.
We also need to note that there can be a difference in phase between elements in $\SUND{2}{1}({\phi},\theta,0)$ and $\DWD_2 (\tilde{z}, \tilde{x})$, therefore, a factor of $\ui$ may be necessary to equate the two.
However, the complex behaviour of characteristic functions is well-known, and it is instead standard practice to define the characteristic as the absolute value or absolute value squared, where such practice is commonplace in the signal processing community, for example when analyzing the analog of the Weyl function -- the ambiguity function.\Cite{Nadav2004, Howard2006}

This allows us to relate the two approaches to a Weyl function by taking the absolute value squared of each or by adding an extra phase, where
\begin{equation}\label{DiscreteContinuousWeylComparison}
	|\DWeyl(\tilde{z},\tilde{x})|^2  = | \chi_{A}(\tilde{z}\pi/2,\tilde{x}\pi/2,0)|^2,
\end{equation}
it's also interesting to see the Weyl function as a generalized autocorrelation function, where for a pure state $\chi_{A}(\phi,\theta,\Phi) = \bra{\psi}\ \SUND{2}{1}(\phi,\theta,\Phi) \ket{\psi}$.
We can think of the rotation operator as acting on the ket state and then taking the inner product.
It then makes sense to take the absolute value squared to get a real-valued autocorrelation function.

Note that in \Eq{DiscreteContinuousWeylComparison}, we relate the continuous Weyl function to the discrete case by setting $\Phi = 0$, this is the simplest way to reduce to two degrees of freedom for easy calculation and equivalent results.
If, on the other hand, we chose to set $\Phi = -\phi$, like we do when equating with the Heisenberg-Weyl case, there would be no way to produce the $\Sz$ operator, hence the lack of informational completeness in this slice for the eigenstates of $\Sz$ in \Eq{informationalIncompleteness}.

The informational completeness for \Eq{DiscreteQubitWeyl} can be seen by observing
\begin{equation}\label{ReverseWeylDiscrete}
	A = \frac{1}{2} \sum_{\tilde{z},\tilde{x}} \DWeyl(\tilde{z},\tilde{x}) \DWD_2 (\tilde{z}, \tilde{x}),
\end{equation}
which demonstrates the power of the use of the Weyl function as a tool for practical purposes in state reconstruction.
For example, the discrete form of the Weyl function has proven to be useful in verification protocols,\Cite{FlammiaLiu} where the Weyl function can be used as a probability distribution for choice of measurement bases.
In the case of stabilizer states,\Cite{GottesmanPhD} this results in the Weyl function having values of $0$ over the majority of the distribution and $1$ when the Weyl kernel is the stabilizer for that state.
This is simply seen by noting that a stabilizer state is an eigenstate of a product of Pauli operators.
For a single qubit, this result is simply seen by noting
\begin{equation}
	\ket{\psi} = \DWD_2(z,x)\ket{\psi} \implies \IP  {\psi}{\psi} = 1 = \bra{\psi} \DWD_2(z,x)\ket{\psi} = \Trace{\rho\,  \DWD_2(z,x)},
\end{equation}
the same naturally holds when extending to multiple qubits. 
Note, however, that in \Refe{FlammiaLiu} the Weyl function is normalized by a factor of $1/\sqrt{2^n}$, for $n$ qubits.
This is done so the factor of $1/2$ is not needed in \Eq{ReverseWeylDiscrete} and $\sum_{\tilde{z},\tilde{x}} \DWeyl(\tilde{z},\tilde{x})^2 = 1$.

Having defined an appropriate displacement operator that generates an informationally complete Weyl characteristic function, to complete the discrete formulation in terms of a generalized displaced parity operator, it is necessary to generate the generalized parity operator. 
The generalized parity is, from \Eq{Pi00}, simply
\begin{equation}
	\DWPi_2 = \DWPi_2(0,0) = \frac{1}{2}\left( \Bid + \Sz   + \Sy + \Sy   \right)
\end{equation}
and
\begin{equation}
	\DWPi_2(z,x) = \DWD_2(z,x)\, \DWPi_2 \, \DWD_2^\dagger(z,x),
\end{equation}
is the kernel written in displaced parity form.

Interestingly, the eigenvalues for $\DWPi_2$ are the same as those for $\SUNPi{2}{1}$, namely $(1\pm\sqrt{3})/2$.
This means that the non-diagonal Wootters generalized parity can be diagonalized as 
\begin{equation}\label{DPDWootters}
	\DWPi_2 = \SUND{2}{1}(\vartheta,\varphi) \, \SUNPi{2}{1} \SUND{2}{1}(\vartheta,\varphi)^\dagger, 
\end{equation}
where,
\begin{equation}\label{KernelStrat2Wootters}
	\vartheta = \arccos{\frac{1}{\sqrt{3}}} \hspace{1cm} \text{and} \hspace{1cm} \varphi = -\frac{\pi}{4}.
\end{equation}
In fact, like the Weyl function, the full Wootters kernel can be written in terms of the Stratonovich kernel, where
\begin{equation}\label{WoottersStratonovichEquivalence}
	\DWPi_2(z,x) = \SUNPi{2}{1}\left(\vartheta+x\pi, \varphi  + \frac{2z-x}{2}\pi\right),
\end{equation}
therefore, the Wootters kernel can be considered a subset of the Stratonovich kernel, and the Wootters qubit Wigner function is a sub-quasidistribution function of the Stratonovich qubit Wigner function.

Note that the four points on the Stratonovich qubit Wigner function where the Wootters function lies form a tetrahedron within the Bloch sphere, as can be seen in \Fig{QubitExamples}, coinciding with the formulation of symmetric informationally complete projective measurements.\Cite{RenesSIC2004,Appleby2005}
We can transform between the Wigner and Weyl functions here by performing the discrete Fourier transform
\begin{equation}
	\DWig(z,x) = \frac{1}{2}\sum_{\tilde{z},\tilde{x}} \DWeyl(\tilde{z},\tilde{x}) \exp\left(\ui \pi\left[ \tilde{x}z - x\tilde{z} \right] \right)
\end{equation}
and
\begin{equation}
	\DWeyl(\tilde{z},\tilde{x}) = \frac{1}{d}\sum_{z,x} \DWig(z,x)\exp\left( - \ui \pi\left[ \tilde{x}z - x\tilde{z} \right] \right).
\end{equation}
The Fourier transform between the Wigner and Weyl function can therefore be seen as the transformation between Pauli basis measurements and symmetric informationally complete projective measurements.

From \Eq{DPDWootters} and \Eq{KernelStrat2Wootters}, where we consider the Wootters kernel as a subset of the full Stratonovich kernel, we can define $Q$ and $P$ functions in discrete phase space.
By using the parity operators from \Eq{QubitQPParity} we can define the general kernel
\begin{equation}
	\DWPi_2^{(s)} = \SUND{2}{1}(\vartheta,\varphi) \, \SUNPi{2}{1(s)} \SUND{2}{1}(\vartheta,\varphi)^\dagger, 
\end{equation}
that yields
\begin{equation}
	\DWPi_2^{(s)} = \DWPi_2^{(s)}(0,0) = \frac{1}{2}\Bid +  \frac{3^{\frac{s}{2}}}{2} \left(\Sz   + \Sy + \Sy   \right).
\end{equation}
For the $Q$ function, this results in a set of symmetric informationally complete positive operator-valued measures.\Cite{RenesSIC2004}
Considering this in terms of frame theory in quantum measurement,\Cite{Ferrie2008,Ferrie2009} consider this discrete $Q$ function as a frame, where the dual frame is then the $P$ function kernel, such that
\begin{eqnarray}
	M &=& \DWPi_2^{(-1)} = \frac{1}{2}\Bid +  \frac{1}{2\sqrt{3}} \left(\Sz   + \Sy + \Sy   \right) \\
	\tilde{M} &=& 3\DWPi_2^{(-1)} - \Bid = \frac{1}{2}\Bid +  \frac{\sqrt{3}}{2} \left(\Sz   + \Sy + \Sy   \right) = \DWPi_2^{(1)}.
\end{eqnarray}
More generally we can transform between any of these kernels by
\begin{equation}\label{DiscreteQubitTransform}
	\DWPi_2^{(s_2)} = 3^{\frac{1}{2}(s2-s1)}\DWPi_2^{(s_1)} + \frac{1-3^{\frac{1}{2}(s2-s1)} }{2} \Bid.
\end{equation}
By understanding these phase-space functions in these terms, they can prove to be powerful tools for verification of quantum states, for example~\Refe{Kliesch2020} which we will discuss further in \Sec{QTinPS}.
	
The relation between the Stratonovich and Wootters formulation of phase-space functions for qubits is a special case that doesn't generalize as the dimension increases.
This is because of the difference in geometry between the representations.
Following this discussion, we will see two variations in how a qudit can be represented on a continuous phase space. 
One representation keeps the \SU{2} group structure and is the natural extension of the Stratonovich kernel, that was also introduced in his seminal paper.\Cite{Stratonovich56}
This method creates a discrete Wigner function on a sphere, where the radius is proportional to the dimension of the Hilbert space.
The second continuous method uses the structure of \SU{d}, where $d$ is the dimension of the Hilbert space.
Geometrically, this is an oblate spheroid. 

The Wootters function, on the other hand, is defined on $\mathcal{Z}(d)\times\mathcal{Z}(d)$ discrete toroidal lattice, where $d$ is prime.
There have been many formulations to extend this form of the Wigner function to higher dimensions, the original Wootters formulation was defined with respect to fields and so the dimensions size was restricted to primes.
Wootter then extended this formulation with Gibbons \textit{et al} to work for dimensions that are powers of primes.\Cite{PhysRevA.70.062101}
Later work also considered the extension to odd numbered dimensions,\Cite{Cohendet_1988, Ruzzi_2005, Gross2016}, and even numbered dimensions.\Cite{Leonhardt1995}
Given the different choices, we refer the reader to any of these articles for specifics, or to \Refe{Gross2016, Ferrie2009, VOURDAS1997367} for more detailed overview and comparison.
Since this construction above the qubit case isn't important to this review, the explicit forms will not be given.

\subsection{SU(2), spin-j}

We will now show the generalization of continuous phase space for larger finite systems.
Here we begin by staying within the \SU{2} Lie group structure before going on to the \SU{N} formulation in \Sec{SUNWigner}.
Both cases can be used to consider similar situations, but also have some subtle differences.
The \SU{2} formulation requires another variable $M = 2j$, two times the azimuthal quantum number, that allows us to represent a spin-$j$ qudit where the dimension of the Hilbert space is $d = M+1$.

It can alternatively be used to model an $M$-qubit state in the symmetric subspace, and can be used over composite methods for certain processes, such as single-axis twisting or Tavis-Cummings interactions.\Cite{Agarwal1997,TavisCummings,Dicke1954,EverittCatt}
These methods can further be used on states that aren't fully symmetric, by taking the approach from Zeier, Glaser, and colleagues\Cite{Garon2015, Leiner2017, Leiner2018,Leiner2020} a multi-qubit state can be represented by a collection of \SU{2} Wigner functions, $n \geq M \geq 0$ for $n$ qubits, to show the Wigner function for every $M$-valued symmetric-subspace.
This leads to many different Wigner functions as the number of qubits increases.

We begin by considering the general \SU{2} rotation operator.
Like with the qubit case, there are many ways of formulating rotations on a sphere.
The extension to the two ways given in \Eq{SU2QubitRotations} are simply
\begin{equation}\label{SU2Rotations}
	\SUND{2}{M}(\phi,\theta,\Phi) = \exp\left(\ui\SUNJ{2}{M}{3}\phi\right)\exp\left(\ui\theta\SUNJ{2}{M}{2}\right)\exp\left(\ui\Phi\SUNJ{2}{M}{3}\right),\hspace{0.5cm} \text{and } \hspace{0.5cm} \OpR_j(\xi) = \exp\left(\xi\ \OpJp_j - \xi^* \OpJm_j\right),
\end{equation}
where
\begin{eqnarray}
	\SUNJ{2}{M}{1} &=& \OpJx_j = \frac{1}{2} \sum _{m=-j}^j \sqrt{j(j+1)-m(m+1)} \left(\ket{j,m}\bra{j,m+1} + \ket{j,m+1}\bra{j,m}    \right)\\
	\SUNJ{2}{M}{2} &=& \OpJy_j = \frac{\ui}{2} \sum _{m=-j}^j \sqrt{j(j+1)-m(m+1)} \left(\ket{j,m}\bra{j,m+1} - \ket{j,m+1}\bra{j,m}\right) \\
	\SUNJ{2}{M}{3} &=& \OpJz_j = \sum _{m=-j}^j  m \ket{j,m}\bra{j,m}
\end{eqnarray}
are the the angular momentum operators, the higher-dimensional extension of the Pauli operators in \SU{2}, and $\OpJpm_j = \SUNJ{2}{M}{1} \pm \ui \SUNJ{2}{M}{2}$ are the angular momentum raising and lowering operators.
Here we also introduce the generalized \SU{N} operator notation $\SUNJ{N}{M}{i}$.

The intuition from looking at $R_j(\xi)$ is that as $j \rightarrow \infty$, the Heisenberg-Weyl displacement operator from \Eq{HWDisplacement} is recovered.
This result was shown by Arecchi \textit{et al.}~in \Refe{Arecchi1972}.
The procedure to show this was also demonstrated in \Refe{Amiet1}, where first we need a contraction of the operators, where we take
\begin{equation}
	\OpApm_j = c\OpJpm_j, \hspace{1cm} \Az = \frac{1}{2c^2}\Bid_j - \SUNJ{2}{M}{3}
\end{equation}
such that, when $c\rightarrow 0$
\begin{equation}
	\lim_{c\rightarrow 0}\OpAp_j = \Opad, \hspace{0.5cm} \lim_{c\rightarrow 0} \OpAm_j = \Opa, \hspace{0.5cm}\text{and},\hspace{0.5cm} \lim_{c\rightarrow 0} \Az = \Opad\Opa.
\end{equation}
The rotation operator therefore becomes
\begin{equation}
	\OpR_j(\xi) = \exp\!\left(\xi \OpAp /c - \xi^*\OpAm/c\right),
\end{equation}
by setting 
\begin{equation}
	\lim_{c \rightarrow 0} \frac{\xi}{c} = \alpha \hspace{0.5cm} \lim_{c\rightarrow 0} \frac{\xi^*}{c} = \alpha^*,
\end{equation}
we yield
\begin{equation}
	\lim_{c \rightarrow 0} \OpR_j(\xi) = \HWD(\alpha).
\end{equation}

A similar treatment can be applied to the Euler angles by noting that $\SUNJ{2}{M}{2} = \ui (\OpJp_j - \OpJm_j)$.
This leads to the result 
\begin{equation}
\lim_{c \rightarrow 0} \SUND{2}{M}(-\phi,-\theta,-\Phi) = \left[\lim_{c \rightarrow 0} \ue{\ui(\phi+\Phi)/2c^2}\right] \exp\left(-|\alpha|^2\right)\exp\left( \alpha\Opad \right)\exp\left( -\alpha^*\Opa \right)   \exp\left((\phi+\Phi)\Opad\Opa \right)
\end{equation}
that, when taking $\Phi = -\phi$ the limit on the right goes to 1, yielding an equivalence to $\HWD(\alpha)$.
Further, when applying this to the vacuum state, we yield
\begin{eqnarray}
\begin{split}
	\lim_{c \rightarrow 0} \SUND{2}{M}(-\phi,-\theta,-\Phi)\ket{j,j} &= \left[\lim_{c \rightarrow 0} \ue{\ui(\phi+\Phi)/2c^2}\right]\exp\left(-|\alpha|^2\right)\exp\left( \alpha\Opad \right)\exp\left( -\alpha^*\Opa \right)   \exp\left(\ui(\phi+\Phi)\Opad\Opa \right)  \ket{0}\\
	&= \left[\lim_{c \rightarrow 0} \ue{\ui(\phi+\Phi)/2c^2}\right] \exp\left(-|\alpha|^2\right)\exp\left( \alpha\Opad \right)\exp\left( -\alpha^*\Opa \right) \exp\left(\ui\phi+\ui\Phi\right) \ket{0}\\
	&= \left[\lim_{c \rightarrow 0} \ue{\ui(\phi+\Phi)/2c^2}\right]\ue{\ui\phi+\ui\Phi-|\alpha|^2/2} \sum_{n=0}^{\infty} \frac{\alpha^n}{\sqrt{n!}}\ket{n},
\end{split}
\end{eqnarray}
equivalent to the coherent state generated by the standard displacement operator.
Where the $\phi$ and $\Phi$ degrees of freedom act as a global phase and cancel out when we set $\Phi=-\phi$.
Note also in this result that the limit of the highest weighted spin state is the vacuum state.

By analogy, the definition of a generalized spin coherent state is just the highest weight state rotated by a suitable operator.
Either rotation operator will work, however they may differ by a phase factor.
For simplicity, we define the spin coherent state by
\begin{equation}
	\ket{(\theta,\phi)_j} = \OpR_j(\theta,\phi)\ket{j,j}.
\end{equation}
This results in the simple definition of the $Q$ and $P$ functions with respect to the spin coherent states
\begin{equation}\label{spinQP}
	Q_A(\theta,\phi) = \bra{(\theta,\phi)_j} \OpA \ket{(\theta,\phi)_j}, \hspace{0.5cm} \text{and,} \hspace{0.5cm} \OpA = \int_0^{2\pi}\int_0^\pi P_A(\theta,\phi)\, \ket{(\theta,\phi)_j}\!\bra{(\theta,\phi)_j}\sin\theta\,\ud\theta\,\ud\phi,
\end{equation}
where it has been shown in \Refe{Koczor2020,KoczorPhD} that these two functions also reach the Heisenberg-Weyl $Q$ and $P$ functions for $j \rightarrow \infty$.

The general idea for the infinite limit of these functions is to imagine a plane tangent to the north pole of the qubit's sphere.
As the value of $j$ increases, the radius of the sphere correspondingly increases, coming closer and closer to the tangential plane at the north pole.
At the limit, the sphere completely touches the plane.

This visualization exercise also helps to understand the difference in Weyl function between \SU{2} and the Heisenberg-Weyl group, where the general \SU{2} Weyl function is
\begin{equation}
	\chi_A(\tilde{\phi},\tilde{\theta},\tilde{\Phi}) = \Trace{\OpA \, \SUND{2}{M}(\tilde{\phi},\tilde{\theta},\tilde{\Phi})}
\end{equation}
As mentioned in the single qubit case, the general \SU{2} Weyl function requires all three Euler angles.
This can be seen in a variety of ways, most importantly it is because when using $\OpR_j(\theta,\phi)$ as the kernel, the resulting function for $\ket{j,j}$ and $\ket{j,-j}$ are identical.
Furthermore, any mixed state on the line inside the Bloch sphere between these two states are also identical.
To resolve this, a third degree of freedom is needed, resulting in the preference for the Euler angles.
This is not a problem for the Heisenberg-Weyl case as any state on the equator is mapped to infinity, and only the equivalent of the top hemisphere is considered in the phase space.
It is also worth noting that in the qubit case, the Weyl function is still informationally complete when $\Phi = 0$, this is not the case with a general qudit.
As we saw when considering the Wootters kernel, there's a special connection between the qubit on the sphere and the qubit on a torus, this means that for the Weyl function, setting $\Phi = 0$, we are modelling the qubit state on a continuous torus.
However, due to how the $\su{2}$ algebra scales as as $M$ increases, the state can no longer be considered on a torus and the $\Phi$ degree of freedom is necessary to generate an informationally complete function.

Next we need to show how the Wigner function can be generated.
Since the original formulation of the general \SU{2} Wigner function, there have been many variations in how it is presented.
From a harmonic series approach to presenting it as a kernel with which a group action can be taken.
Since all of these representations are equivalent, we will present how the kernel can be generated through a multipole expansion.

For this, we need a new set of generators to generalize the Pauli operators from the qubit case.
These are known as the tensor or Fano multipole operators,\Cite{Fano53}
\begin{equation}
	\OpT^j_{lm} = \sqrt{\frac{2l+1}{2j+1}} \sum_{m',n = -j}^j C^{jn}_{jm',lm}\ket{j,n}\bra{j,m'},
\end{equation}
where $C^{jn}_{jm',lm}$ are Clebsch-Gordan coefficients that couple two representations of spin j and l to a total spin j.
The kernel can then be generated from there operators by
\begin{equation}\label{KlimovKernel}
	\SUNPi{2}{M}(\theta,\phi) = \sqrt{\frac{\pi}{M+1}}\sum_{l=0}^{M}\sum_{m=-l}^l Y_{lm}^*(\theta,\phi) \OpT_{lm}^{M/2},
\end{equation}
where $Y_{lm}^*(\theta,\phi)$ are the conjugated spherical harmonics.
By following the same procedure from \Eq{Pi00}, the \SU{2} generalized parity can be derived
\begin{equation}\label{SU2MParity}
	\SUNPi{2}{M} = \SUNPi{2}{M}(0,0) = \sum_{l=0}^{M}\frac{2l+1}{M+1} \sum_{n= -j}^{j}C^{jn}_{jn,l0}\ket{j,n}\bra{j,n},
\end{equation}
where we note again that $j = M/2$.
Note that when taking the limit $j \rightarrow \infty$, this generalized parity operator becomes the parity operator defined in \Eq{HWParity}. 
The derivation and proof of this can be found in \Refe{Amiet1}.
Resulting in the generalized displaced parity operator
\begin{equation}\label{SU2SSS}
	\SUNPi{2}{M}(\theta,\phi) = \SUND{2}{M}(\phi,\theta,\Phi) \SUNPi{2}{M} \SUND{2}{M}(\phi,\theta,\Phi)^\dagger,
\end{equation}
where, since the $\Phi$ degrees of freedom cancel out with the diagonal generalized parity, the full kernel is equivalent to the Heisenberg-Weyl kernel from \Eq{HWKernel} as $j \rightarrow \infty$. \cite{Amiet1}

More generally, the $Q$ and $P$ functions from \Eq{spinQP}, and any $s$-parameterized quasi-probability distribution function, can be generated by a simple extension of \Eq{KlimovKernel} and \Eq{SU2MParity}, where the general kernel can be calculated\Cite{BrifMann1999,Koczor2020}
\begin{equation}\label{su2sValuedKernel}
	\SUNPi{2}{M(s)}(\theta,\phi) = \sqrt{\frac{\pi}{M+1}}\sum_{l=0}^M (C^{jj}_{jj,l0})^{-s}\sum_{m=-l}^l Y_{lm}^*(\theta,\phi) \OpT_{lm}^{M/2},
\end{equation}
where the extra value $(C^{jj}_{jj,l0})^{-s}$ is 1 when $s=0$, returning \Eq{KlimovKernel}.
The generalized parity for any $s$-valued function can then be calculated
\begin{equation}
	\SUNPi{2}{M(s)} = \SUNPi{2}{M(s)}(0,0) = \sum_{l=0}^{M}\frac{2l+1}{M+1} \left(C^{jj}_{jj,l0} \right)^{-s}\sum_{n= -j}^{j}C^{jn}_{jn,l0}\ket{j,n}\bra{j,n},
\end{equation}
which for $s = -1$ yields
\begin{equation}
	\SUNPi{2}{M(-1)} = \ket{j,j} \bra{j,j},
\end{equation}
Where, as mentioned earlier, these functions also become their infinite-dimensional counterparts as $j \rightarrow \infty$.\Cite{Koczor2020,KoczorPhD}

From \Eq{GenFourierTransform}, we can then relate and transform between these different quasi-distribution functions by way of a generalized Fourier transform.
However, there is an alternative method by way of spherical convolution\Cite{Koczor2020}.
By using the highest-weight state in the Wigner representation and performing a convolution using \Eq{GenConvultion} we can transform an $s$-valued function to the corresponding $(s-1)$-valued function for the same state.
For example, we can transform from the $P$ function to the Wigner function, and the Wigner function to the $Q$ function by using this method.
Similarly, we can transform from the $P$ function of a state to the corresponding $Q$ function by using the convolution with the $Q$ function highest weighted state.

\subsection{SU(N)}\label{SUNWigner}

We now consider an alternative route to treat a general $d$-level or multi-qubit quantum system.
Like the \SU{2} case, an \SU{N} representation can also be used for a general qudit state, where in this case $d = N$ and we only consider $M=1$. 
In order to represent an $n$-qubit state, the construction of an $\SU{2^d}$ kernel is required -- it is also possible to take the same approach as with $\SU{2}$ and consider only the symmetric subspace with an \SU{d} representation.
Alternatively, the generalized displacements and generalized parities can be mixed and matched to some extent as was shown in \Refe{TilmaEveritt}, this case will also be considered here.

Although we provide a formalism to construct these Wigner functions, the details to actually construct the kernels can prove to be complicated, there are therefore some open questions for generalized \SU{N} kernels.
the first open problem is generalizing \SU{N} for any value of $M$, effectively creating the analog of the symmetric subspace for $M$ $N$-level qudits.
However, the procedure to create the generalized displacement operators for this case can be found in \cite{TilmaKae1}.

The second open problem is the consideration of the hyperbolic space of \SU{Q,P}.
A construction of an \SU{1,1} Wigner function has recently been provided in \Refs{Seyfarth,Klimov2020}, however the general \SU{Q,P} case remains to be solved.
Insight into these cases may lead to some interesting results in ideas relating to AdS/CFT in a phase space formalism.
Note that if one is also interested in some of the more fantastical routes to the unification of quantum mechanics and gravity, in theory a construction of $E_8$ should be possible, however unwieldy the number of degrees of freedom prove to be.
Here we will settle ourselves with providing the results from the \SU{1,1} phase space construction to close this section.

The kernel for an \SU{N} phase space now requires a specific generalized displacement and generalized parity.
The first step is to construct the generators of \SU{N}, which can be represented by $N^2-1$ matrices of size $N\times N$.
The method to construct these can be found in \Refe{Nemoto2000}.
This results in a set of operators $\SUNJ{N}{1}{i}$, where $1 \leq i \leq N^2-1$.
In fact, only a subset of the generators are needed to construct the \SU{N} Euler angles, these are generalizations of the \Sy and \Sz operators, which are
\begin{eqnarray}
	\mathtt{J}^{\mathsf{y}}_N(1,a) &=& \SUNJ{N}{1}{(a-1)^2+1} = \ui \left( \ket{a}\bra{1} - \ket{1}\bra{a} \right),\\
	\mathtt{J}^{\mathsf{z}}_N(a^2-1) &=& \SUNJ{N}{1}{a^2-1} = \sqrt{\frac{2}{a(a-1)}}\left(\sum_{i = 1}^{a-1} \ket{i}\bra{i}  \right) - \sqrt{\frac{2(a-1)}{a}}\ket{a}\bra{a}, \label{cartanSubSUN}
\end{eqnarray}
where $2\leq a \leq N$.
Note that this is a subset of the full set of generator of \su{N}; $N^2-1$ elements are needed to generate \su{N} however just $2(N-1)$ elements of the algebra are needed to generate the generalized Euler angles for \SU{N}.
We also note that there has been an alternative construction of phase space for \SU{3} where they instead used three dipole and five quadrupole operators as the generators of the Lie algebra\Cite{Hamley2012}.

Following this, there is a procedure to generate the generalization of the Euler angles for \SU{N}, that can be found in \Refs{Tilma2,Tilma3,Nemoto2000}.
This results in the generalized Euler angles $\SUND{N}{1}(\boldsymbol{\phi},\boldsymbol{\theta},\boldsymbol{\Phi})$, where $\boldsymbol{\phi} = \{\phi_1, ..., \phi_{N(N-1)/2} \}$, $\boldsymbol{\theta} = \{\theta_1, ..., \theta_{N(N-1)/2} \}$ and $\boldsymbol{\Phi} = \{\Phi_1, ..., \Phi_{N-1} \}$.
This rotation operator can be split up into two parts, first one dependent on the $\boldsymbol{\phi}$ and $\boldsymbol{\theta}$ degrees of freedom, $\hat{\mathcal{R}}_N^1(\boldsymbol{\phi},\boldsymbol{\theta})$, and the second dependent on the $\boldsymbol{\Phi}$ degrees of freedom, $\hat{\mathcal{S}}_N^1(\boldsymbol{\Phi})$, such that
\begin{equation}
	 \hat{\mathcal{R}}_N^1(\boldsymbol{\phi},\boldsymbol{\theta}) \equiv  \prod_{N\geq a \geq 2} \prod_{2\leq b \leq a} \exp\left(\ui \mathtt{J}^\mathsf{z}_N(3)\phi_{a-1+k(b)} \right)\exp\left( \ui \mathtt{J}^\mathsf{y}_{N}(1,a)\theta_{a-1+k(b)} \right),
\end{equation}
where
\begin{equation}
	k(b) = \begin{cases}
		0,\hspace{1.35cm} b = N,\\ 
		\sum_{i=1}^{N-b}, \hspace{0.5cm} b\neq N,
	\end{cases}
\end{equation}
and
\begin{equation}
	\hat{\mathcal{S}}_N^1(\boldsymbol{\Phi}) \equiv \prod_{1\leq c \leq N-1} \exp\left( \ui \mathtt{J}^\mathsf{z}_{N}([c+1]^2)\Phi_{N(N-1)/2+c}\right).
\end{equation}
Finally yielding the \SU{N} rotation operator
\begin{equation}\label{SUNRotation}
	\SUND{N}{1}(\boldsymbol{\phi},\boldsymbol{\theta},\boldsymbol{\Phi}) = \hat{\mathcal{R}}_N^1(\boldsymbol{\phi},\boldsymbol{\theta})\hat{\mathcal{S}}_N^1(\boldsymbol{\Phi}). 
\end{equation}

Just from this generalization of the Euler angles, many of the results from the \SU{2} case can be extended to \SU{N}.
Most straight-forward is an informationally complete Weyl characteristic function that takes the form
\begin{equation}
	\chi_A(\tilde{\boldsymbol{\phi}},\tilde{\boldsymbol{\theta}},\tilde{\boldsymbol{\Phi}}) = \Trace{\OpA \, \SUND{N}{1}(\tilde{\boldsymbol{\phi}},\tilde{\boldsymbol{\theta}},\tilde{\boldsymbol{\Phi}})}.
\end{equation}
Note that a discrete alternative can be yielded by constructing all $N^2-1$ generators for \SU{N}, resulting in
\begin{equation}
	\chi_A(k) = \Trace{\OpA \, \SUNJ{N}{1}{k}}
\end{equation}
where $0\leq k \leq N^2-1$, $\SUNJ{N}{1}{0}=\Bid_N$ and the other $\SUNJ{N}{1}{k}$ matrices are the generators of the \su{N} algebra.

Also from the rotation operator in \Eq{SUNRotation} one can generate generalized coherent states.\Cite{Nemoto2000}
Like earlier examples of coherent states, \SU{N} coherent states are generated by applying the rotation operator to a suitable analog to the vacuum state.
In a similar procedure \SU{N} coherent states can be generated by the $\ket{0}_N^1$ state, where we define this state as being the eigenstate of $\lambda_N(N^2-1)$ with the lowest eigenvalue, $-[2(N-1)/N]^\frac{1}{2}$, making it the lowest-weighted state, resulting in 
\begin{equation}
	\ket{(\boldsymbol{\phi},\boldsymbol{\theta})^1_N} = \SUND{N}{1}(\boldsymbol{\phi},\boldsymbol{\theta},\boldsymbol{\Phi}) \ket{0}_N^1 = \begin{pmatrix}
		\ue{\ui (\phi_1 + \phi_2 + ... + \phi_{N-1})} \cos(\theta_1)\cos(\theta_2)...\cos(\theta_{N-2})\sin(\theta_{N-1})\\
		-\ue{\ui (-\phi_1 + \phi_2 + ... + \phi_{N-1})} \sin(\theta_1)\cos(\theta_2)...\cos(\theta_{N-2})\sin(\theta_{N-1})\\
		-\ue{\ui (\phi_3 + \phi_3 + ... + \phi_{N-1})} \sin(\theta_2)\cos(\theta_3)...\cos(\theta_{N-2})\sin(\theta_{N-1})\\
		\vdots\\
		-\ue{\ui (\phi_{N-2}+\phi_{N-1})} \sin(\theta_{N-3})\cos(\theta_{N-2})\sin(\theta_{N-1})\\
		-\ue{\ui (\phi_{N-1})} \sin(\theta_{N-2})\sin(\theta_{N-1})\\
		\cos(\theta_{N-1})
	\end{pmatrix},
\end{equation}
ignoring an overall global phase.
We note that it is also possible to set the analog to the vacuum state as the eigenstates of $\SUNJ{N}{1}{3}$ with eigenvalue 1, as is the more standard choice for the use of \SU{2} for qubits.
However, given the structure of the generators for \su{N}, taking the highest-weight state results in much tidier calculations.

From the construction of a coherent state, the generations of the $Q$ and $P$ functions are simply
\begin{equation}
	Q_A(\boldsymbol{\phi},\boldsymbol{\theta}) = \bra{(\boldsymbol{\phi},\boldsymbol{\theta})^1_N} A \ket{(\boldsymbol{\phi},\boldsymbol{\theta})^1_N}, \hspace{0.5cm} \text{and } \hspace{0.5cm} A = \int_{\boldsymbol{\Omega}} P(\boldsymbol{\phi},\boldsymbol{\theta})\ket{(\boldsymbol{\phi},\boldsymbol{\theta})^1_N}\bra{(\boldsymbol{\phi},\boldsymbol{\theta})^1_N} \ud\boldsymbol{\Omega},
\end{equation}
where $\boldsymbol{\Omega}$ are the degrees of freedom on \SU{N} and $\ud \boldsymbol{\Omega}$ in the volume normalized differential element for \SU{N}, the construction of which can be found in \Refe{Tilma2}.
Similarly, the kernel for the \SU{N} Wigner function can be generated with respect to coherent states, where\Cite{TilmaKae1}
\begin{equation}\label{WignerSUNKernel}
	\Pi_N^1(\boldsymbol{\phi},\boldsymbol{\theta}) = \frac{1}{N}\Bid_N + \frac{\sqrt{N+1}}{2} \sum_{k=1}^{N^2-1} \bra{(\boldsymbol{\phi},\boldsymbol{\theta})^1_N} \SUNJ{N}{1}{k} \ket{(\boldsymbol{\phi},\boldsymbol{\theta})^1_N} \SUNJ{N}{1}{k}
\end{equation}

By setting the variable to 0, the generalized parity operator for \SU{N} is yielded, where
\begin{equation}\label{SUNParity}
	\SUNPi{N}{1} = \frac{1}{N}\left( \Bid_N - \sqrt{\frac{(N-1)N(N+1)}{2}}\mathtt{J}_z^N(N^2-1) \right),
\end{equation}
producing an operator constructed by the identity operator and the diagonal operator from \Eq{cartanSubSUN}, where $a=N$.
Like the construction of an \SU{2} generalized parity operator, the \SU{N} parity is  a diagonal operator and commutes with the $\hat{\mathcal{S}}_N^1(\boldsymbol{\Phi})$ part of the rotation operator, resulting in the degrees of freedom cancelling out.
The number of degrees of freedom are further reduced to allow the \SU{N} Wigner function to live on the complex projective space $\CP{N-1}$ with $2(N-1)$ degrees of freedom.
By considering the operators, this can be realized by noting that each \SU{N} generator acts on $N$ \SU{2} subspaces of the full \SU{N}, where each subspace is acted on by $\mathtt{J}_y^N(1,a)$.
Since $N-1$ elements of the generalized parity operator are the same, with just the last element differing, it is in effect a weighted identity operator for the rotations that do not act on the last \SU{2} subspace.
And so, the elements of the rotation operator that have as the generator $\mathtt{J}_y^N(1,a)$ for $a<N$ cancel out, until we reach the element that rotates around $\mathtt{J}_y^N(1,N)$.

Note that if we stay with the fundamental representation of \SU{N} there is a simple transformation between the different quasiprobability distribution functions.
This can be shown by noting that the generalization of \Eq{WignerSUNKernel} is\Cite{TilmaKae1}
\begin{equation}
	\Pi_N^{1(s)}(\boldsymbol{\phi},\boldsymbol{\theta}) = \frac{1}{N}\Bid_N + \frac{(N+1)^{\frac{1}{2}(1+s)}}{2} \sum_{k=1}^{N^2+1} \bra{(\boldsymbol{\phi},\boldsymbol{\theta})^1_N} \SUNJ{N}{1}{k} \ket{(\boldsymbol{\phi},\boldsymbol{\theta})^1_N} \SUNJ{N}{1}{k},
\end{equation}
resulting in the general function
\begin{equation}
	F_{N,1,\DO }^{(s)}(\boldsymbol{\phi},\boldsymbol{\theta}) = \frac{1}{N} + \frac{(N+1)^{\frac{1}{2}(1+s)}}{4} \sum_{k=1}^{N^2+1} \bra{(\boldsymbol{\phi},\boldsymbol{\theta})^1_N} \SUNJ{N}{1}{k} \ket{(\boldsymbol{\phi},\boldsymbol{\theta})^1_N} \Trace{\DO\SUNJ{N}{1}{k}}.
\end{equation}
We can then simply transform between functions by taking
\begin{equation}\label{SUNTransform}
	F_{N,1,\DO }^{(s_1)}(\boldsymbol{\phi},\boldsymbol{\theta}) = \frac{1}{N} + (N+1)^{\frac{1}{2}(s_1-s_2)} \left(F_{N,1,\DO }^{(s_2)}(\boldsymbol{\phi},\boldsymbol{\theta})-\frac{1}{N}\right),
\end{equation}
providing a general formula to consider qudits in \SU{N} similar to \Eq{DiscreteQubitTransform} for qubits.
Note in the same way as the qubit case, one can define a set of symmetric informationally complete measurements from the $Q$ function, this can then be transformed into its dual by 
\begin{equation}
	\Pi_N^{1(1)}(\boldsymbol{\phi},\boldsymbol{\theta}) = (N+1)\Pi_N^{1(-1)}(\boldsymbol{\phi},\boldsymbol{\theta}) - \Bid,
\end{equation}
which corresponds to the transform between an $N$-dimensional frame and its dual frame.\Cite{Ferrie2009,Ferrie2008}

This all provides a structure for considering qudits in phase space, alternatively by considering the \SU{2} subgroup structure of \SU{N}, we can provide a link between the different representations of multi-qubit states.
It also means that we can alternatively represent multi-qubit states with a hybrid approach of $\SU{2}$ and $\SU{N}$ that was introduced in \Refe{TilmaEveritt} and applied in \Refe{Rundle2016}.
From applying the procedure from \Eq{CompositeKernel} to generate a kernel for composite systems, for multi-qubit states we get
\begin{eqnarray}\label{CompositeSUN}
\begin{split}
	\SUNPi{2^{\otimes n}}{1}(\boldsymbol{\theta},\boldsymbol{\phi}) &= \bigotimes_{i=1}^n \SUNPi{2}{1}(\theta_i,\phi_i), \\
	&= \left(\bigotimes_{i=1}^n \SUND{2}{1}(\phi_i,\theta_i,\Phi_i)\right) \left(\bigotimes_{i=1}^n \SUNPi{2}{1} \right) \left(\bigotimes_{i=1}^n  \SUND{2}{1}(\phi_i,\theta_i,\Phi_i)\right)^\dagger,\\
	&= \nQubitD{n}(\boldsymbol{\phi},\boldsymbol{\theta},\boldsymbol{\Phi})\,\left(\bigotimes_{i=1}^n \SUNPi{2}{1} \right)\,\nQubitD{n}^\dagger(\boldsymbol{\phi},\boldsymbol{\theta},\boldsymbol{\Phi}),
\end{split}
\end{eqnarray}
where $\nQubitD{n}(\boldsymbol{\phi},\boldsymbol{\theta},\boldsymbol{\Phi})$ is the $n$-qubit rotation operator.

We can take this new $n$-qubit rotation operator and apply it to the $\SU{N}$ generalized parity operator to create an alternative representation of a multi-qubit state in phase space
\begin{equation}
	\SUNPi{2^{[n]}}{1}(\boldsymbol{\theta},\boldsymbol{\phi}) = \nQubitD{n}(\boldsymbol{\phi},\boldsymbol{\theta},\boldsymbol{\Phi})\,\SUNPi{N}{1} \,\nQubitD{n}^\dagger(\boldsymbol{\phi},\boldsymbol{\theta},\boldsymbol{\Phi}),
\end{equation}
that also obeys the Stratonovich-Weyl correspondence S-W.~\ref{SW1}~-~\ref{SW5}.
Each choice of representing qubit states has its own benefits and drawbacks.
We can immediately see that the benefit of using the multi-qubit \SU{2} rotation operator with the \SU{N} generalized parity operator is the reduced number of degrees of freedom.
The multi-qubit rotation operator results in $2n$ degrees of freedom, whereas the \SU{N} operator will result in $2(2^n -1)$ degrees of freedom for $n$ qubits.

While on the topic of the \SU{N} representation of quantum states, it's also worth noting a slightly different formulation; $\SU{P,Q}$, that has the same number of degrees of freedom as \SU{P+Q}.
However, the geometry of \SU{P,Q} differs significantly from the compact \SU{N} geometry, resulting from a difference in sign between the first $P$ elements and the remaining $Q$ elements in the metric, yielding a non-compact, hyperbolic geometry.
The case of a generalized \SU{P,Q} representation in phase space is still an open question.
However the specific case of \SU{1,1} has been addressed in a recent paper.\Cite{Seyfarth}
The \SU{1,1} representation of certain quantum systems has proven to be useful in simplifying some particular problems in quantum mechanics.
In particular Hamiltonians involving squeezing.
This is because the $\SU{1,1}$ displacement operator is a squeezing operator, resulting in the \SU{1,1} coherent states to actually be squeezed states.
Since the squeezing of quantum states leads to many important results in quantum technologies, such as metrology in particular, the use of these phase-space methods for squeezing can prove to be a useful tool.

The generators of algebra of $\SU{1,1}$ are similar to $\SU{2}$ with a difference in the commutation relations, where
\begin{equation}
	[\OpKx^M,\OpKy^M] = -\ui\OpKz^M, \hspace{0.5cm} [\OpKy^M,\OpKz^M] = \ui\OpKx^M, \hspace{0.5cm} [\OpKz^M,\OpKx^M] = \ui\OpKy^M.
\end{equation}
The raising and lowering operators are then constructed 
\begin{equation}
	\OpKpm^M = \pm \ui \left( \OpKx^M \pm \OpKy^M \right),
\end{equation}
that, along with $\OpKz^M$, constitutes an alternative basis, where the generalized displacement operator is
\begin{equation}\label{SU11Rotation}
	\SUPQD{M}(\zeta) = \exp\left(\OpKp^M\zeta - \OpKm^M\zeta^* \right)
\end{equation}
much like \Eq{SU2Rotations}, where $\zeta$ is the complex squeezing parameter.
The central result from \Refe{Seyfarth} the generalized displaced parity operator for \SU{1,1} is constructed from \Eq{SU11Rotation} and the \SU{1,1} generalized parity $(-1)^{\OpKz^M}$, yielding
\begin{equation}
	\SUNPi{1,1}{M}(\zeta) = \SUPQD{M}(\zeta)\,(-1)^{\OpKz^M}\, \SUPQD{M}^\dagger(\zeta),
\end{equation}
as the kernel for $\SU{1,1}$.

Some of the authors from \Refe{Seyfarth} also looked at the equivalence between $\SU{2}$ and $\SO{3}$ in phase space~.
Such analysis could also be used to go between $\SU{1,1}$ and $\SO{1,2}$ in phase space, or even generalized to $\SO{1,3}$ or higher, to consider relativistic quantum mechanics in phase space.
There have been works that have looked at generalizing the Wigner function for relativistic particles, for instance see \Refs{Carinena1990, Hakim1978, Shin1992, Sirera1999}.
Although the results are interesting and potentially useful, it is beyond the scope of this paper to discuss them here.

\section{Quantum technologies in phase space}\label{QTinPS}

Given the mathematical framework, we will now look into how the representation of quantum systems in phase space can be used to understand processes in quantum technologies.
We will give some examples of experiments that have taken place to directly measure the phase-space distribution of quantum states.
This will include both discrete-variable and continuous-variable systems, as well as the hybridization of the two.
We will also mention ideas that haven't yet been put into practice, but are worth discussion.

Before the practical results, we will consider how some important metrics, already present in the quantum information community, can be described within the phase-space formulation.
One metric that is unique to phase-space functions is the presence of `negative probabilities'.
Due to the uniqueness and unintuitive nature of such a phenomena, this will receive a full discussion shortly, where we will discuss what negative values do and do not say about a quantum state.
Before looking at negative volume of the Wigner function as a metric, we will consider other figures of merit that can be calculated through phase-space methods

\subsection{Measures of quality}\label{Metrics}

There are measures that are widely used when considering the density matrix or state vector of a quantum state that can be expressed within the phase-space representation of quantum mechanics. 
In particular, considering the Wigner function and a key property of traciality from \upshape{S-W.}~\ref{SW4}, where
\begin{equation}\label{traceWigner}
	\Trace{\OpA\OpB} = \int_\Omega W_A(\Omega)W_B(\Omega)\, \ud\Omega,
\end{equation}
many similarities naturally fall out.
The most natural starting point is to consider the expectation value of some observable $\OpA$ with respect to some state $\rho$ such that
\begin{equation}
	\left\langle \OpA \right\rangle_\rho = \Trace{\rho\OpA} = \int_\Omega W_\rho(\Omega)W_A(\Omega)\, \ud\Omega.
\end{equation}
In the case of a qudit in \SU{2}, we note that the expectation value of the angular momentum operators can be related to the center of mass for the corresponding Wigner function, where
\begin{equation}
	\left\langle \SUNJ{2}{M}{i} \right\rangle_\rho = \frac{(2j+1)\sqrt{j(j+1)}}{8\pi}\int_0^{2\pi}\int_0^\pi f_i(\theta,\phi)W_\rho(\theta,\phi) \sin\theta\, \ud\theta\,\ud\phi,
\end{equation}
where\Cite{Chen2019}
\begin{equation}\label{WignerHarmonics}
	f_i(\theta,\phi) = \{ -\sin\theta\cos\phi,-\sin\theta\sin\phi,\cos\theta\}, \hspace{0.5cm} \text{since} \hspace{0.5cm} W_{\SUNJ{2}{M}{i}}(\Omega) = \sqrt{j(j+1)}f_i(\Omega)
\end{equation}
where the minus signs and normalization come from our formulation of the Wigner function kernel, which may differ in sign is following a different convention.

There are further important properties in quantum information that utilize the trace of two operators, as such their construction in the phase-space representation are straightforward.
A simple example of this is the purity of a state, which for density matrices is $\mathsf{P}(\rho) = \Trace{\rho^2}$, can be calculated in phase space by
\begin{equation}\label{Purity}
	\mathsf{P}[W_\rho(\Omega)] =  \int_\Omega W_\rho(\Omega)^2 \, \ud\Omega.
\end{equation}
Next we consider the fidelity of a state $\rho_2$ with reference to a target state $\rho_1$, where 
\begin{equation}
	\mathsf{F}(\rho_1,\rho_2) = \Trace{\sqrt{\sqrt{\rho_1}\rho_2\sqrt{\rho_1}}}^2.
\end{equation}
We are often interested in comparing an experimentally generated state $\rho_2$ with a pure target state $\rho_1$, in which case the fidelity reduces to $\mathsf{F}(\rho_1,\rho_2) = \Trace{\rho_1\rho_2}$, which when applied to \Eq{traceWigner} we simply yield
\begin{equation}\label{WignerFidelity1}
	\mathsf{F}\left[ W_{\rho_1}(\Omega), W_{\rho_2}(\Omega) \right] = \int_\Omega W_{\rho_1}(\Omega)W_{\rho_2}(\Omega)\, \ud\Omega.
\end{equation}
The phase-space calculation of fidelity leads to an interesting method of fidelity estimation of measured quantum states. 
We will provide the procedure how one can do that here, following the work on Flammia and Liu.\Cite{FlammiaLiu}

First, it is important to note that the fidelity can be calculated using any valid distribution $F(\Omega)$, including the Weyl function and the $Q$ and $P$ functions, so long as we have the corresponding dual function $\tilde{F}(\Omega)$ resulting in
\begin{equation}\label{WignerFidelity}
	\mathsf{F}\left[ \rho_1, \rho_2 \right] = \int_\Omega \tilde{F}_{\rho_1}(\Omega)F_{\rho_2}(\Omega)\, \ud\Omega,
\end{equation}
where in this sense the Wigner function is self-dual, and the $P$ function is the dual to the $Q$ function, and \textit{vice versa}.
One drawback from the structure of \Eq{WignerFidelity} is that we need to integrate over the full phase space, however we know from \Sec{FiniteSystems} that only four points in the phase space are required to produce an informationally complete function.
Therefore for $n$ qubits we require a sum over $2^n$ points in phase space.
We can then calculate fidelity through the sum
\begin{equation}\label{DiscretePSFidelity}
	\mathsf{F}(k) = \sum_k F_{\rho_2}(k)\tilde{F}_{\rho_1}(k).
\end{equation}
Note that for this to hold, and for later calculations, we also require the functions to be normalized such that $\sum_k (F_{\rho_1}(k))^2 =\sum_k (\tilde{F}_{\rho_1}(k))^2 = 1$.

Following \Refe{FlammiaLiu}, we then need to create an estimator for the fidelity.
For this, select any measurement $k$ with probability $\mathrm{Pr}(k) = (\tilde{F}_{\rho_1}(k))^2$.
The estimator is then 
\begin{equation}
	X = \frac{F_{\rho_2}(k)}{\tilde{F}_{\rho_1}(k)}, \hspace{0.5cm} \text{such that} \hspace{0.5cm} E[X] = \sum_k (\tilde{F}_{\rho_1}(k))^2 \frac{F_{\rho_2}(k)}{\tilde{F}_{\rho_1}(k)} = \Trace{{\rho_1}{\rho_2}},
\end{equation}
where $E[X]$ is the expected value of $X$.
When measuring in practice, calculating the function $F_{\rho_2}(k)$ perfectly is not possible, and so we need to make many copies of ${\rho_2}$ and perform sufficient measurements to assure a given confidence of fidelity.
The details of which can be found in \Refe{FlammiaLiu}.

Here we would like to note that besides being applicable to any number of qubits, such a procedure can also be used for any number of qudits, where we can use the \su{N} algebra to generate a qudit Weyl function or a set of informationally complete measurements on the manifold of pure states in \SU{N}.

Fidelity is a measure of how close a measured state is to some target state. 
We may conversely be interested in a distance measure, that measures how far away (in Hilbert space) a given state is from some target state.
Such a distance measure is known as the trace distance, where 
\begin{equation}\label{TD}
	d_T(\rho_1,\rho_2) = \frac{1}{2}\lVert \rho_1-\rho_2\rVert_1 = \frac{1}{2}\Trace{\sqrt{(\rho_1-\rho_2)^\dagger(\rho_1-\rho_2)}},
\end{equation}
where $1-\sqrt{\mathsf{F}(\rho_1,\rho_2)} \leq d_T(\rho_1,\rho_2) \leq \sqrt{1 - \mathsf{F}(\rho_1,\rho_2)}$.

For a qubit, \Eq{TD} can instead be expressed in terms of the Wigner function, where
\begin{equation}
	d_T(\rho_1,\rho_2) = \left( \int_\Omega\left[W_{\rho_1}(\Omega)-W_{\rho_2}(\Omega)\right]^2 \ud\Omega \right)^{\frac{1}{2}} = \left( \int_\Omega\left[W_{\rho_1-\rho_2}(\Omega)\right]^2 \ud\Omega \right)^{\frac{1}{2}}
\end{equation}

This can be seen by noting the two states can be expressed as
\begin{equation}
	\rho_1 = \frac{1}{2}\Bid  + \frac{1}{2}\sum_i a_i \Si{i}, \hspace{0.5cm} \rho_2 = \frac{1}{2}\Bid  + \frac{1}{2}\sum_i b_i \Si{i}, \hspace{0.25cm} \text{and so}\hspace{0.25cm} \rho_1-\rho_2 = \frac{1}{2}\sum_{i} (a_i-b_i) \Si{i} 
\end{equation}
resulting in the trace distance being
\begin{equation}
	d_T(\rho_1,\rho_2) = \frac{1}{4} \Trace{\left(\sum_{i} (a_i-b_i)^2\Bid \right)^{\frac{1}{2}}} = \frac{1}{2} \left(\sum_{i} (a_i-b_i)^2 \right)^{\frac{1}{2}}.
\end{equation}
Similarly the Wigner function can be expressed
\begin{equation}
	W_{\rho_1-\rho_2}(\Omega) = \frac{\sqrt{3}}{2}\sum_i (a_i-b_i) f_i(\theta,\phi)
\end{equation}
from \Eq{WignerSUNKernel} along with the properties in \Eq{WignerHarmonics}.
This results in
\begin{equation}
	\int_\Omega [W_{\rho_1-\rho_2}(\Omega)]^2 \, \ud\Omega = \frac{3}{4}\int_\Omega \left(\sum_{i,j} (a_i-b_i)(a_j-b_j) f_i(\theta,\phi)f_j(\theta,\phi) \right) \, \ud\Omega = \frac{1}{4} \sum_{i} (a_i-b_i)^2  
\end{equation}
therefore
\begin{equation}
	\left(\int_\Omega [W_{\rho_1-\rho_2}(\Omega)]^2 \, \ud\Omega \right)^\frac{1}{2} = \frac{1}{2} \left( \sum_{i} (a_i-b_i)^2  \right)^\frac{1}{2} = d_T(\rho_1,\rho_2).
\end{equation}
Such a procedure can then be generalized to multi-qubit states.

Further, similarly to the calculation of fidelity, we can exchange the integral over the continuous degrees of freedom by performing a sum over discrete degrees of freedom. 
Where the procedure can be applied to not only the Wigner function, but the all distributions, albeit with a different normalization out the front.

We also note that alternatively, one can find the Monge distance with respect to the $Q$ function.\Cite{Zyczkowski1998}
This effectively calculated the distance between the centers of mass of two distributions.
Given the non-negative property of the Q function, it can be useful in translating procedures performed on classical probability distributions into quantum mechanics.
As such, it can also be used in calculating the R\'enyi entropy.
The R\'enyi entropy is defined with respect to the Q function as\Cite{Renyi1961,Wehrl1978}
\begin{equation}\label{RenyiEntropy}
	S_R = -\int_\Omega Q_\rho(\Omega) \log Q_\rho(\Omega) \, \ud\Omega.
\end{equation}
This result was extended to Wigner functions in \Refe{PhysRevA.51.2575}, however this is limited to the Heisenberg-Weyl case.
We note here that in the case of \SU{N}, one can simply use \Eq{SUNTransform} to write \Eq{RenyiEntropy} in terms of the Wigner function.

The above set of equations provide a framework that show how the phase-space representation of quantum mechanics can prove a viable alternative to the more traditional density operator approach.
All of the above can also be extended for any choice of phase-space function, such as the $Q$, $P$, and Weyl functions.
And where the equations were specifically states for qudit states, general expressions can be found to calculate these values for \SU{N} structures of even continuous-variable systems. 
Further to these measures there are other ways in which phase-space methods have been useful to characterize a quantum state, these are: measures of coherence;\Cite{Sperling2018, Sperling2020} phase-space inequalities;\Cite{park2020,Bohmann2020,Bohmann2020-2,Biagi2021}; and the negativity in the Wigner function, which plays an important role.
The manifestation of these negative values provide nuanced details for each system.
A closer look into negative values will be considered next.

\subsection{Negative probabilities}\label{WigNeg}

Negative probabilities have been a subject of interest since the early days of quantum mechanics.
From the observation of negative values in the Wigner function, Dirac later discussed their presence along with negative energies in 1942\Cite{DiracNegProb}. 
Dirac stated that negative probabilities `should not be considered as nonsense. They are well-defined concepts mathematically, like a negative of money.'
This growing acceptance to the concept of negative probabilities then lead to a number of other people to take the concept seriously.
A more rigorous exploration was undertaken two years later by Bartlett.\Cite{bartlett_1945}

Later, Groenewold argued that to satisfy other desirable properties of a quantum-mechanical probability distribution function in phase space, it is necessary for negative probabilities to exist.\Cite{Groenewold1946}
In the '80s, Feynman then gave a simple argument for the existence of negative probabilities in constructing a discrete Wigner function.\Cite{Feynman1987-FEYNP}.
Put simply, his argument was that negative probabilities are no more nonsense than a negative quantity of anything (in his example he gave apples), and that the negativity is just one step in a total sum.
The negativity is never considered in isolation.

Similarly, any probability over phase space needs to be averaged over a quadrature, due to the inherent uncertainty of quantum states.
When a measurement is made over a quadrature, the resulting probability is always non-negative.
The Wigner function is unique among different phase-space distributions as it results in the correct marginals, see \Eq{marginals}.

But, what does the negativity actually tell us when it appears in the Wigner function?
It is generally accepted that it is a result of quantum correlations, or quantum interference.
By putting two coherent states in a superposition, the coherences from the two states will interfere creating oscillating positive a negative interference in between, that we interpret as quantum correlations.
This has led to results where the negative volume of a Wigner distribution was calculated, where
\begin{equation}
	V[W_\rho(\Omega)] = \frac{1}{2}\left( \int_\Omega \left| W_\rho(\Omega) \right|\ud\Omega  - 1\right),
\end{equation}
which can be seen as an additional measure of quality derived from the Wigner function.
Such a measure can be used for any continuous distrbution, and even composite systems.\Cite{Arkhipov2018}

The full story is that this negativity is actually a consequence of non-Gaussianity, a result that is known as Hudson's theorem.\Cite{HUDSON}
Consequently negativity doesn't capture the important quantum correlations that arise from squeezing, in particular two-mode squeezing.
Which, since the state is a squeezed Gaussian, has no negative quasi-probabilities.
So some care needs to be taken in using negative volume as a measurement of quantumness, non-classicality, entanglement and so on.
That being said, it doesn't mean negativity shouldn't be used, and there have been many results showing its use.\cite{Kenfack2004, Lutkenhaus1995, Siyouri2016, Taghiabadi2016}
It just may be more important to look at the context in some cases, and look for how correlations manifest in particular ways.

In the same way, the interpretation of negative values in discrete quantum systems needs some care.
Any coherent, single-qubit state generated by the Stratonovich kernel will exhibit negative values.
The maximum positive value is in the direction that the Bloch vector is pointing, as we deviate from that point over phase space, the probability slowly decreases to zero and then negative, where the maximal negative value is at the point orthogonal to the Bloch vector.
Because of the term `classical states' for coherent states, this tends to lead to the confusion that coherent qubits states, such as the standard choice of basis in quantum information $\ket{\uparrow}$ and $\ket{\downarrow}$, are classical.
This term `classical' is just a misnomer, a qubit state is fundamentally non-classical, likewise with any quantum coherent state.
By definition, any coherent qubit state is quantum; the only non-quantum state that a qubit can exist in is the fully mixed state.
Negativity can therefore be used as a sign of quantumness, but it's not necessarily a sign of entanglement.

\begin{figure}
	\includegraphics[width=\linewidth]{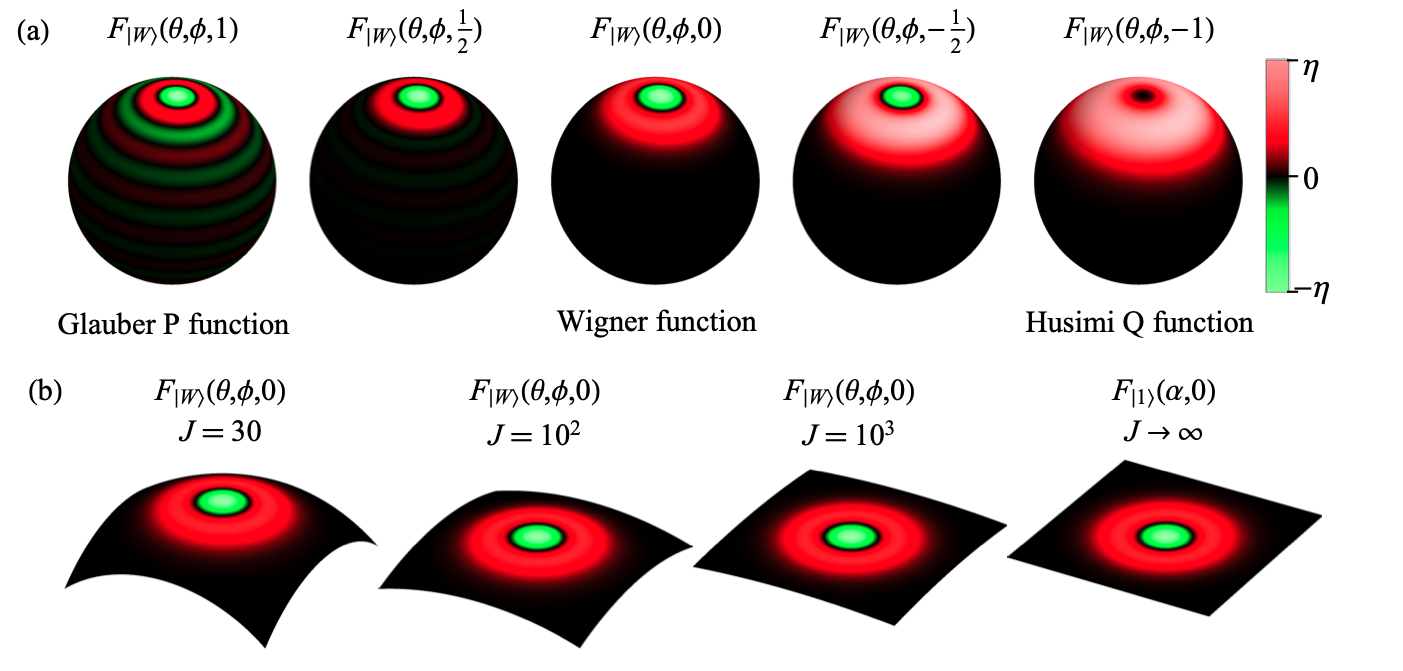}
	\caption{\label{jInfinityFock} Examples of the phase space distribution of the Dicke state $\ket{j,j-1}$, also known as the W state if considered as the symmetric state of $2j$ qubits, from \Refe{Koczor2020}.
	Note that the $J$ in the figure corresponds to our $j$ in the text.
	\Sub{a} shows the Dicke state for $j=10$ where $F_{\ket{W}}(\theta,\phi,s)$ shows that this is the s-valued phase-space function for the W state.
	As with \Eq{su2sValuedKernel} $s=0$ corresponds to the Wigner function and $s=\pm1$ refers to the P and Q functions respectively.
	Intermediate values of $s$ have also been taken in between.
	\Sub{b} shows the Wigner function where the value of $j$ increases from $j=30$ up to the limit $j\rightarrow \infty$, showing that the W state tends towards the 1-photon Fock state as $j\rightarrow \infty$.
	Reprinted figure with permission.\textsuperscript{\cite{Koczor2020}} Copyright (2020) by the American Physical Society.}
\end{figure}

One could also make the argument that this negativity can be considered a sign of superposition, as $\ket{0}$ is a superposition of $(\ket{0}+\ket{1})/\sqrt{2}$ and $(\ket{0}-\ket{1})/\sqrt{2}$, and in fact any pure state for a qubit is the superposition of two antipodal pure states.
In this way, the negativity is a result of this and the overall symmetry of qubit states; this really shows the beauty of the Stratonovich kernel, how it presents the symmetry in coherent qudit states, where every $N$-level coherent state is just the \SU{N} rotation of the chosen ground state.

If one decides that this symmetry is not so important and require that specific spin coherent states are non-negative, then they can consider the Wootters kernel instead.
The main benefit to the Wootters kernel is that the eigenstates of the Pauli matrices all produce non-negative Wigner functions.
In fact, if an octahedron is drawn within the Bloch sphere, where the vertices are the eigenstates of the Pauli matrices, the discrete Wigner function for every state that lies on and within this octahedron is non-negative.
These are the so-called magic states,\Cite{Howard2014} if a pure state rotates between any of these magic states, then negativity arises -- a result that is made clear by considering the Wootters kernel as a subset of the Stratonovich kernel.
This gives rise to the literature on the equivalence between Wigner function negativity and contextuality, and further work on magic state distillation, proving to be an advantage to choose the Wootters kernel over the Stratonovich kernel.\cite{Spekkens2008,Delfosse2015,Delfosse_2017,Raussendorf2017,Schmid2018,Schmid2020,Schmid2021}
The negativity that arises from using Wootters kernel can further be summarized by considering the extension of Hudson's theorem for finite-dimension systems produced by Gross.\Cite{Gross2016}

Alternatively, when using the Stratonovich kernel, one could consider an analog of Hudson's theorem to apply to the decreasing volume of negativity in the highest weighted state, $\ket{j,j}$, as $j\rightarrow\infty$. 
At the infinite limit the state becomes Gaussian and the Heisenberg-Weyl group is reached -- the highest weighted state is simply the non-negative vacuum state.
In the same way, by choosing the Dicke state with one energy level lower $\ket{j,j-1}$, as $j\rightarrow\infty$ this slowly transforms into the one-photon Fock state, as is shown in \textbf{\Fig{jInfinityFock}}.
In \FigSub{jInfinityFock}{a} we see the phase-space representation of the Dicke state $\ket{j,j-1}$ where $j=10$; this also shows how the state is represented in phase space for different values of $s$, including giving a good demonstration of how negative values arise for different parameterizations.
We then focus on the Wigner function in\FigSub{jInfinityFock}{b}, where we can see how this Dicke state goes to the 1-photon Fock state at the limit $j\rightarrow\infty$.
Similar results can also be seen from taking the limit $j\rightarrow\infty$ for any Dicke state, where $\ket{j,j-n}\rightarrow \ket{n}$, the $n$-photon Fock state.


\subsection{Methods for optical systems}


Phase-space methods have gained much attention in the optics community, where there is a large collection of books on quantum optics that include in-depth treatment of phase-space methods and the application thereof.
This will therefore not be a thorough account of the advancements made in quantum optics using these methods, we will instead just point out some results we find particularly interesting and will prove useful later in this section.
For anyone who wants to read more deeply into the application of phase-space methods in quantum optics we suggest the following heavily truncated list of books on quantum optics.\Cite{Schleich, agarwal_2012, Carmichael, GerryKnight, Scully1997}
There have been further, in-depth reviews of the use of the Wigner function in optical systems\Cite{Curtright2014, Ferry} and in the use of optical quantum computing\Cite{RevModPhys.77.513,PhysRevA.65.062309}.

When generating the Wigner function experimentally, there are many approaches one can take. 
The two we are interested in here are first the generation of the density operator and then following \Eq{WignerFunctionHW} to produce the Wigner function.
The second is to take direct measurements of points in phase space, by passing the need to calculate the full density operator.
The real power of measuring quantum systems in phase space comes in realizing the displaced parity formalism of the kernel.
By realizing the invariance of cyclic permutations of elements within a trace operation, it is clear that the Wigner function generation can be seen as the parity measurement of a displaced state:
\begin{equation}\label{HWMeasurement}
	W_\rho(\alpha) = \Trace{\DO\,\HWPi(\alpha) } = \Trace{\DO\,\HWD(\alpha)\HWPi \HWD^\dagger(\alpha)} = \Trace{\HWD^\dagger(\alpha) \DO\HWD(\alpha)\,\HWPi } = \Trace{\DO(-\alpha)\,\HWPi },
\end{equation}
where $\HWD(-\alpha) = \HWD^\dagger(\alpha)$, and $\DO(\alpha) = \HWD(\alpha) \DO \HWD(\alpha)^\dagger$ is the rotated density operator.

This insight amounts to an experimental procedure of creating the quantum state, then displacing the state by some value $\alpha$ and then measuring this state.
At this point the parity operator can be applied through classical computation of diagonal elements, or by following the procedure in~\Refe{PhysRevLett.78.2547}, one can perform a parity measurement by coupling the optical state to a two-level atom, where the two levels act as the $\pm1$ parity values.

Direct measurement of the Weyl function, on the other hand, hasn't gained so much attention.
Similarly to the measurement of the Wigner function, this can be executed by coupling to a qubit, where instead of taking parity measurements, the qubit is measured in the $x$ and $y$ Pauli basis.
Following the procedure laid out in \Refs{Tufarelli2011,Tufarelli2012}, we start by coupling the optical state, $\rho_f$ with a qubit in the state $\ket{+} = (\ket{\uparrow}+\ket{\downarrow})/\sqrt{2}$, such that
\begin{equation}
	\rho_\text{tot}(0) = \ket{+}\bra{+}\otimes\rho_f.
\end{equation}
We then perform the unitary transformation
\begin{equation}
	\rho_\text{tot}(\tilde\alpha) = \mathsf{R}(\tilde\alpha) \rho_\text{tot}(0) \mathsf{R}^\dagger(\tilde\alpha), \hspace{1cm}\text{where}\hspace{1cm} \mathsf{R}(\tilde\alpha) = \exp\left(\frac{1}{2} \Sz \otimes \left[ \Opad\tilde\alpha - \Opa \tilde\alpha^* \right] \right),
\end{equation}
note that this is the Pauli $z$ operator with the displacement operator with a factor of a half.
Once the state has been prepared in the appropriate point in phase space, $\alpha$, the qubit is then measured in either the $x$ or $y$ basis, resulting in
\begin{equation}\label{WeylMeas}
	\HWWeyl_\rho(\HWWeylvar) = \langle \Sx \rangle + \ui \langle \Sy \rangle,
\end{equation}
where $\langle \Sx \rangle = \Trace{\Sx \rho_\text{tot}(\tilde\alpha)}$ and $ \langle \Sy \rangle = \Trace{\Sy   \rho_\text{tot}(\tilde\alpha)}$.
We note that a similar procedure can be performed by coupling a qubit to a qudit to generate the Weyl characteristic function for qudit states.

Once we have generated the Weyl function, we can use \Eq{HWGeneralPDF} to perform a simple Fourier transformation to generate any quasi-probability distribution function.
An important part of the analysis of optical states is the investigation of different transforms of this Weyl characteristic function, in order to reveal certain quantum effects that are more difficult to see.
One such case where this approach is insightful is when considering two-mode squeezed states.
Such states are Gaussian and so have no negative values, despite being highly entangled.
Generating a characteristic function and then filtering to a quasiprobability can result in negative values.\Cite{PhysRevLett.107.113604,Kok2020}
Allowing a phase-space analysis of such states.\Cite{PhysRevA.86.052118,park2020}
A similar approach has also been taken for analyzing discrete states.\Cite{Kiesel2012}
An experimental way to achieve these results should be possible in combination with an extension of \Eq{WeylMeas}.

\subsection{Measuring qubits in phase space}

By building on the techniques used to measure phase space for continuous variable systems, it is possible to visualize discrete variable systems in phase space.
Some early work sought to first recreate the density operator for the state, before applying the appropriate kernel.
One such example of this method was to show the creation of an atomic Schr\"odinger cat state in an experiment using a Rydberg atom with angular momentum $j=25$.\Cite{Signoles2014}
This work takes the atom as a large angular momentum space, with a Hilbert space of dimension 51.

Alternatively, one can measure a multi-qubit state by considering the collection of symmetric subspaces that the full Hilbert space is made up of, in a tomography method known as DROPS (or discrete representation of operators for spin systems).\Cite{Garon2015,Leiner2017,Leiner2018,Leiner2020}
Other methods would rely on taking the $j=n/2$ symmetric subspace of an $n$-qubit system, the power of using the DROPS method is that you get much more information about the state -- including information about antisymmetric states such as the two-qubit singlet state, which cannot be represented by the $j=1$ symmetric subspace.
Another benefit of this method is that the authors have made vast libraries of results available, including code to generate these functions.\Cite{SpinDrops}

The methods already mentioned rely on the full construction of the density matrix for the state, requiring full tomography before a phase-space representation can be constructed.
As the number of qubits increases, this can get exponentially difficult, we therefore desire methods to represent a quantum state in phase space that don't require full density matrix reconstruction.

Our main focus from now on is the Wigner function and to present results that utilize the displaced parity structure of the kernel, where we can imitate the Heisenberg-Weyl group result from \Eq{HWMeasurement}.
By using the framework outlined in \Sec{FiniteSystems} these results can be adapted into whichever phase-space distribution is preferred, as they are all informationally complete representations of a quantum state.
First we will consider the simplest case of phase-space functions generated by a single \SU{2} kernel construction from \Eq{SU2SSS}.
This will then be followed up with utilizing \Eq{CompositeSUN} to show how a composition of single-qubit kernels can be used to measure the phase space of quantum states directly.

Adapting \Eq{HWMeasurement} for use in general \SU{2} systems results in
\begin{eqnarray}\label{wigMeasurement}
	W_\rho(\theta,\phi) = \Trace{\DO \, \SUNPi{2}{M}(\theta,\phi) } &= \Trace{ \DO \, \SUND{2}{M}(\phi,\theta,\Phi) \SUNPi{2}{M} \SUND{2}{M}(\phi,\theta,\Phi)^\dagger} \nonumber\\
	&= \Trace{ \DO(\phi,\theta,\Phi) \,  \SUNPi{2}{M}} = \left\langle  \SUNPi{2}{M} \right\rangle_{\DO(\phi,\theta,\Phi)}
\end{eqnarray}
where $\DO(\phi,\theta,\Phi) = \SUND{2}{M}(\phi,\theta,\Phi)^\dagger \DO \SUND{2}{M}(\phi,\theta,\Phi)$.
The procedure is to rotate the generated state $\DO$ to a desired point in phase space, creating the state $\DO(\phi,\theta,\Phi)$, followed by taking the appropriate generalized parity measurement for the group structure.

When experimentally measuring phase space the way that \Eq{wigMeasurement} is practically implemented depends on many factors, including the system that's being measured and how the measurement results are collected.
This means that practically, the rotation and generalized parity measurement needs to be put in the right context.
Two examples of this when measuring a single qubit can be found in recent works, \Refe{Tian2018} and \Refe{Chen2019}, where in \Refe{Tian2018} the authors measured a two-level caesium atom, 
and in \Refe{Chen2019} authors measure a state produced in a nitrogen vacancy center in diamond.
In both cases, they first produced a state in a superposition of the excited and the ground state.

Using \Eq{wigMeasurement} by applying pulses to rotate this state by values of $\theta$ and $\phi$ to given point in phase space.
Projective measurements are then taken at each point in phase space, resulting in the spin projection probabilities $p_m(\theta,\phi)$ for $m = \pm 1/2$. 
These projective measurements can be expressed as
\begin{equation}
	p_{\pm\frac{1}{2}}(\theta,\phi) = \Trace{\DO\, \SUND{2}{1}(\theta,\phi)P_{\pm\frac{1}{2}}\SUND{2}{1}(\theta,\phi)^\dagger}
\end{equation}
where the $P_{\pm\frac{1}{2}}$ are the projectors of the eigenstates of $\Sz/2$ with eigenvalues $\pm 1/2$ respectively.
These can also be thought of as the diagonal entries of the density matrix $\DO(\phi,\theta,0)$ from \Eq{wigMeasurement}.

The generalized parity is then applied to the spin projection probabilities by calculating
\begin{eqnarray}\label{QubitTomography}
	W(\theta,\phi) &= \sum_{m=-1/2}^{1/2}p_m(\theta,\phi) \left[\SUNPi{2}{1} \right]_{mm},\nonumber\\
	&= \frac{1}{2}\left[1 + \sqrt{3}\left( p_{\frac12}(\theta,\phi) - p_{-\frac12}(\theta,\phi) \right) \right]
\end{eqnarray}
where $\left[\SUNPi{2}{M} \right]_{mm}$ is the $m^{\text{th}}$ diagonal entry of the generalized parity operator.
Note that here we're using the same normalization of the Wigner function used throughout the text, where as the results in \Refe{Tian2018} and \Refe{Chen2019} differ by a factor of $2\pi$.

\begin{figure}
\begin{center}
	\includegraphics[width = \linewidth]{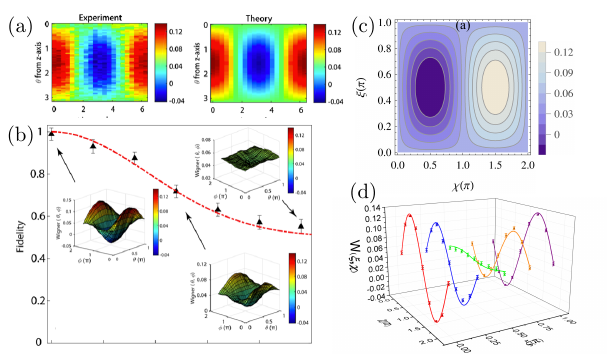}
\end{center}
\caption{\label{SingleQubitExperiments} Examples of phase-space measurement performed on a single qubit.
\Sub{a} and \Sub{b} are results and figures from \Refe{Chen2019}, \Sub{c} and \Sub{d} are similarly from \Refe{Tian2018}.
In both cases, the Wigner function for a single-qubit state has been constructed, where \Sub{a} and \Sub{b} have created an eigenstate of $\Sy$.
\Sub{a} compares the results from experimental and theoretical construction.
\Sub{b} shows the same state as it decoheres, plotting the fidelity over time, with snapshots of the Wigner function at three points of decoherence.
\Sub{c} and \Sub{d} show an eigenstate of $\Sx$, where similarly \Sub{c} shows the theoretical plot for this state.
\Sub{d} shows the experimental results of points in phase space. 
The dots with error bars are the experimental results, where the curves are the theoretical values, showing good comparison between the experimental and theoretical values.
\Sub{a} and \Sub{b} Reproduced with permission.\textsuperscript{\cite{Chen2019}} Copyright (2019), AIP Publishing.
\Sub{c} and \Sub{d} Reprinted figure with permission.\textsuperscript{\cite{Tian2018}} Copyright (2018) by the American Physical Society.
}
\end{figure}

Results from the two experiments are shown in \textbf{\Fig{SingleQubitExperiments}}, where \Fig{SingleQubitExperiments}~\Sub{a} and~\Sub{b} show the results from \Refe{Chen2019} and \Fig{SingleQubitExperiments}\Sub{c} and~\Sub{d} are the results from \Refe{Tian2018}. 
In \Refe{Chen2019}, they measured 1200 points in phase space, where the values of $\theta$ are separated by a step size of $\pi/60$ and the values of $\phi$ have a step size of $\pi/10$.
They also demonstrated the consequence of decoherence on the Wigner function for this state, \Fig{SingleQubitExperiments}~\Sub{b} shows how the phase space flattens as the fidelity decreases.
\Fig{SingleQubitExperiments}~\Sub{c} and~\Sub{d} show the second experiment. 
\Fig{SingleQubitExperiments}~\Sub{c} is the theoretical calculation of the Wigner function.
\Fig{SingleQubitExperiments}~\Sub{d} shows the experimental results, shown as dots with error bars, compared to the theoretical curves; showing good agreement between experiment and theory.

Like with the Wigner function for optical systems, phase-space methods prove useful for showing the coherence in qubits.
However, for systems of qubits to be useful for the advancement of quantum technology, it is important to consider multi-qubit systems.
This is where the power of phase-space methods becomes more apparent.

For a single qubit, full tomography isn't that difficult a task, and requires fewer measurements than measuring all of phase space.
However, when the number of qubits increases, the size of the Hilbert space increases exponentially, and so measuring points in phase space becomes more of a viable alternative.

As we have already mentioned, there are two main approaches to represent a multi-qubit state.
We start with the symmetric subspace, where we are limited to considering symmetric states of composite qubits.
In many situations, symmetric states are all that are desired from the output for use in computational settings or metrology.
Examples of such symmetric states are coherent states\Cite{Arecchi1972,Perelomov,PerelomovB}; GHZ states\Cite{GHZ} or, more generally, atomic Schr\"odinger cat states\Cite{Agarwal1997}; W states, or Dicke states\Cite{Dicke1954}; and spin-squeezed states.\Cite{Agarwal1997,Kitagawa1993,Jian2011}

It is therefore important to understand how we can directly measure the phase space of these larger Hilbert spaces, presenting an alternative to traditional tomographical methods.
It has already been noted that the Wootters-kernel Wigner function can be seen as being a sub-distribution of the Wigner function generated with the Stratonovich kernel.
Since both functions are informationally complete, this means that by taking four carefully chosen points in phase space, the full state can be reconstructed.
In fact, these points don't need to be the four that map the Stratonovich kernel to the Wootters kernel, they just need to be appropriately distributed around the phase space.

Similarly, it is possible to reconstruct the full quantum state, and therefore the full phase-space distribution, by taking a subset of phase-point measurements for any value of $j$.
It was shown in \Refe{Koczor2020} that one can similarly reconstruct the full state by measuring a finite number of points.
This can be realized by generalizing \Eq{QubitTomography} to an \SU{2} Hilbert space of any arbitrary dimension, where
\begin{equation}\label{QuditTomography}
	W(\theta,\phi) = \sum_{m=-j}^{j}p_m(\theta,\phi) \left[\SUNPi{2}{M} \right]_{mm},
\end{equation}
note here we've restricted discussion to the Wigner function, but this can easily be generalized to any phase-space function, see \Refe{Koczor2020} for more details.

By using the result from \Eq{KlimovKernel}, that the kernel can be generated from tensor operators and spherical harmonics, points in the phase space can be redefined in terms of a spherical harmonic expansion, with coefficients $c_{lm}$ that correspond to the spherical harmonic $Y_{lm}(\theta,\phi)$.
The full phase-space function can then be reconstructed by calculating $c_{lm}$ from the Wigner function at $(4j+2)^2$ points in phase space, $W(\theta_a,\phi_b)$, where $\theta_a = a\pi/(4j + 2)$ and $\phi_b = 2b\pi/(4j + 2)$.

Note that the $4j + 2$ points are put as a lower bound, and more points may be necessary given the intricacies of the state and phase space being measured.
However, it is also noted in \Refe{Koczor2020} that this is just the case by equally distributing the angles over phase space resulting in bunching at the poles; using other methods to distribute points over the sphere, such as taking a Lebedev grid\Cite{Lebedev1975,Lebedev1976,Lebedev1999} may reduce the number of points in phase space needed for reconstruction.
For example, taking the case where $j=1/2$ would result in 16 points in phase space, where we know from \Eq{WoottersStratonovichEquivalence} the full state can be reconstructed from 4 points in phase space, distributed as a tetrahedron over the surface of a sphere.
It stands to reason that this reduction in points will scale as we increase $j$.
A method to reduce these points for specific states was demonstrated in \Refe{Koczor2020-2}.
We add that since these $(4j+2)^2$ points in phase space constitute an informationally complete set of measurements, one could then adapt \Eq{DiscretePSFidelity} to calculate the fidelity from these points in phase space.

Alternatively, to measure systems of multiple qubits, we can consider a kernel that is the composition of multiple \SU{2} kernels from \Eq{CompositeSUN}.
As was shown in \Eq{CompositeSUN}, measurement of the phase space for such a composite state can be performed by local rotations on each qubit and then taking a generalized parity measurement of the resulting state.
Each qubit then has two degrees of freedom from the local rotations, this then results in a function that has $2n$ degrees of freedom.

In order to make sense out of this many degrees of freedom, it is necessary to choose certain slices of this high-dimensional phase space.
The most natural choice is to take the equal-angle slice, this is where the number of degrees of freedom is reduced to two, $(\theta,\phi)$, where we set every $\theta$ equal, such that $\theta_1=\theta_2=...=\theta_n = \theta$.
Likewise, we set all the $\phi$ degrees of freedom to equal each other.
The reason this is a natural choice is because the resulting phase-space function shares similarities with the symmetric subspace variant.
For example a $j=n/2$ two-component atomic Schr\"odinger cat state is similar to the equal angle slice of an $n$-qubit GHZ state, where in both cases the Wigner function has two coherent states orthogonal to one-another, say at the north and south pole, with $n$ oscillations around the equator between the two coherent states.
These oscillations are the signature of quantum correlations arising from superposition in Schr\"odinger cat state and entanglement in GHZ states.
In fact, there are similarities between any symmetric $n$-qubit state in the equal-angle slice and the corresponding $j=n/2$ function.
It's also worth noting that for an $n$-qubit $Q$ function for a symmetric state, the equal-angle slice is indistinguishable to the $j=n/2$ $Q$ function for the comparable state.

The equal-angle slice for the Wigner function was also shown to be useful in considering the graph-isomorphism problem.\Cite{Mills2019}
It was shown in \Refe{Mills2019} that if you produce two graph states\Cite{Hein2004,Hein2006} from isomorphic graphs, then their equal-angle slices are equivalent, since taking the equal-angle slice can be thought of as an analogy of disregarding qubit, and in this case node, order.
It was further shown in \Refe{Mills2019} that non-isomorphic graphs have distinguishable equal-angle slices.
However the results get more difficult to analyze as the number of qubits increase, as the detail in the Wigner function gets more and more intricate as the size of the Hilbert space increases.

Apart from the equal-angle slice, there are many choices one may prefer to use, depending on whatever characteristic correlations may arise within the system of interest.
And some of these different choices, along with the equal-angle slice, will be considered in more detail here.

\begin{figure}
\begin{center}
	\includegraphics[width = \linewidth]{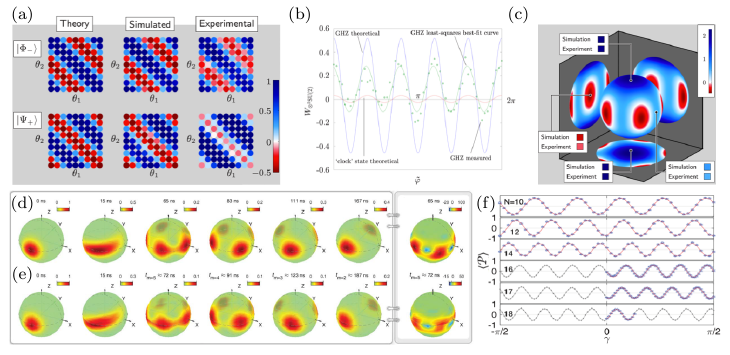}
\end{center}
\caption{\label{MultiQubitExperiments} Examples phase-space measurement performed on multi-qubit systems with a composite kernel calculation.
\Sub{a}, \Sub{b}, and \Sub{c}  are results and figures from \Refe{Rundle2016}, and \Sub{d}~-~\Sub{f} are from \Refe{Song2019}.
In all cases a composite generalized displaced parity form has been taken to measure the phase space of these states.
\Sub{a} shows two two-qubit maximally entangled Bell states, where theoretical plots are shown, followed by simulated and experimental results that are both from IBM's qx.
\Sub{b} and \Sub{c} then show results of measuring a five-qubit GHZ state. 
\Sub{b} shows the equatorial slice -- equally spaced points around the equator of the equal-angle slice.
\Sub{c} shows the full theoretical equal-angle slice with chosen points in phase space measured.
\Sub{d} and \Sub{e} are $Q$ functions of points in a single-axis twisting evolution, where \Sub{d} are theoretical simulations and \Sub{e} are the experimental measurement results.
The resultant states are a spin coherent state, a squeezed spin coherent state, followed by five-, four-, three-, and two-component atomic Schr\"odinger cat states in order from left to right. 
The last panel shows the corresponding Wigner function for the five-component atomic Schr\"odinger cat state.
\Sub{f} shows the equivalent to the equatorial slice from \Sub{b} on $N$-qubit GHZ states where the $\Sz$ operator is used as the generalized parity operator.
 \Sub{a} -~\Sub{c} Reproduced with permission.\textsuperscript{\cite{Rundle2016}} Copyright (2016), AIP Publishing.
\Sub{d} -~\Sub{f} Reproduced with permission.\textsuperscript{\cite{Song2019}} Copyright (2019),The American Association for the Advancement of Science.
}
\end{figure}

Starting with the simplest case of a composite system, we will consider bipartite entanglement between two qubits.
Two qubits famously exhibit entanglement in the well-known maximally entangled Bell states, explicitly
\begin{equation}
	\ket{\Phi_{\pm}} = \frac{1}{\sqrt{2}}\left( \ket{\uparrow\uparrow} \pm \ket{\downarrow\downarrow} \right) \hspace{1cm} \ket{\Psi_{\pm}} = \frac{1}{\sqrt{2}}\left( \ket{\uparrow\downarrow} \pm \ket{\downarrow\uparrow} \right),
\end{equation}
where $\ket{\uparrow\uparrow}$ is shorthand for $\ket{\uparrow}\otimes\ket{\uparrow}$.
Apart from the equal-angle slice, there are other slices one can take to gain an understanding of the quantum correlations present within the system.
One such example was shown in \Refe{Rundle2016}, where the slice $\phi_1=\phi_2=0$ was taken and the resulting $\theta_1$ and $\theta_2$ degrees of freedom were plotted against each other.
An example of this slice for $\ket{\Phi_-}$ and $\ket{\Psi_+}$ from \Refe{Rundle2016} is shown in \textbf{\FigSub{MultiQubitExperiments}{a}}.
Two-qubit Bell-type entanglement in the Wigner function manifests in this slice as oscillations between negative and positive quasi-probabilities in phase space.
\FigSub{MultiQubitExperiments}{a} shows theoretical results for what should be measured at 81 points in phase space for these two states, along with what was actually measured using IBM's qx.

There are further ways to view phase space for bipartite entanglement, for instance the approach in works such as \Refe{Jing2019} consider bipartite entanglement between two states that have larger Hilbert space, demonstrating the entanglement between two Bose-Einstein condensates.
In \Refe{Jing2019}, the authors considered the Wigner function of these states by taking a marginal Wigner function and a conditional Wigner function.
The marginal Wigner function is simply the Wigner function for one of the states, which can be calculated for either the reduced density matrix or by integrating out one set of degrees of freedom
\begin{equation}\label{MarginalWigner}
	W_1(\Omega_1) = \Trace{\DO_1\, \SUNPi{2}{M}(\Omega_1)} =  \int_{\Omega_2} W(\Omega_1,\Omega_2) \,\ud\Omega_2,
\end{equation}
where $\DO_1 = \mathrm{Tr}_2\left[ \DO \right]$ is the reduced density matrix for the first subsystem.
Note that for a maximally entangled Bell states, taking the marginal Wigner function results in a completely mixed state that has a value of $1/2$ everywhere, see the video in the supplementary material of \Refe{Rundle2016} for an example of this.
The conditional Wigner function is calculated by projecting a state onto the basis states of one of the subsystems, more details can be found in \Refe{Jing2019}.

As the number of qudits increases, the choices of slice becomes more difficult, due to the increasing number of degrees of freedom.
In this case, the $n$-qudits can be represented by $n$ marginal distributions, however if there is any entanglement present these become less useful.
As a default, the equal-angle slice is the simplest starting point to getting an idea of the correlations that manifest in phase space.

Examples of the equal-angle slice have been shown to be useful in characterizing multi-qubit states in many experimental settings, where there have been examples ranging from three-qubit states up to 20-qubit multi-component Schr\"odinger cat states.
In \Refe{Ciampini2017}, the authors created entangled states of three qubits, generated by two photonic qubits and a further path-encoded qubit.
Through this method they generated a cluster GHZ state and a W state.

As we move up to a five-qubit state in \FigSub{MultiQubitExperiments}{b} and \FigSub{MultiQubitExperiments}{c}, we can see there are now five oscillations.
When plotting on a sphere, the variation in the $\phi$ degrees of freedom for $\theta = \pi/2$ lies on the equator.
This gives rise to the name `equatorial slice'.
The equatorial slice on it's own has been plotted in \FigSub{MultiQubitExperiments}{b} with both theoretical and experimental results plotted, where there has been a least-squares best-fit curve plotted for the experimental data.
By looking for these oscillations around the equator, this test serves as a verification of the generation of GHZ-type entanglement.
Although there is clearly a lot of noise in the machine, the oscillating behaviour shows that a considerable about of GHZ-type entanglement remains in the state.

\FigSub{MultiQubitExperiments}{b} considers this equatorial slice taking measurement with the Wigner function parity operator.
However, this may not always be the most desirable choice to look for these oscillations, since as the number of qubits increases, the amplitude of the oscillations steadily decreases.
Alternatively, one can try different choices of generalized parity operator, such as in \Refe{Song2019} where $\bigotimes \Sz  $ was taken, as can be seen in \FigSub{MultiQubitExperiments}{f}.
Experimental results for taking the equatorial slice using this parity is shown for 10, 12, 14, 16, 17, and 18 qubits in \FigSub{MultiQubitExperiments}{f}, showing good agreement between theory and experiment.

We note that the use of the \Sz operator as a parity operator does not satisfy the Stratonovich-Weyl correspondence, requiring some extra assumptions for informational completeness, see \Eq{sZParity} and the discussion around it.
However, if we're measuring the state of a quantum system it is safe to make the relevant assumptions -- that the operator is trace 1.
The kernel $\Sz(\theta,\phi) = \SUND{2}{1}(\theta,\phi)\Sz\SUND{2}{1}(\theta,\phi)^\dagger$ can then be considered as a generalized observable, creating a characteristic function that has a Hermitian kernel, where the Pauli operators \Sx, \Sy, and, \Sz are yielded by setting $(\theta, \phi) = (-\pi/2,0)$, $(\theta, \phi) = (-\pi/2,-\pi/2)$, and $(\theta, \phi) = (0,0)$ respectively.
This allows a direct comparison to verification and fidelity estimation protocols, for example \Refs{Guhne2007,Kliesch2020} and the fidelity estimation results in \Sec{Metrics}.

The experiments from \Refe{Song2019} were performed using a 20-qubit machine that generated entangled states by employing a single-axis twisting Hamiltonian.\Cite{Agarwal1997}
The single-axis twisting Hamiltonian also creates different multi-component Schr\"odinger cat states throughout evolution before reaching the bipartite GHZ state.
 \FigsSub{MultiQubitExperiments}{d} and~\Sub{e} show snapshots of different points in the evolution in the $Q$ function representation.
 Showing the Q function for a spin coherent state and a spin squeezed state, followed by a five-, four-, three-, then two-component atomic Schr\"odinger cat state.
 These figures show the power of the $Q$ function to demonstrate the presence of spin coherent states, however it is lacking in the ability to show the level of quantum interferences between states.
In the last panel, the authors show the Wigner function for the five-component Schr\"odinger cat state, where the presence of negative values between each spin coherent state demonstrates the presence of quantum interference and therefore entanglement.

\subsection{Hybrid continuous- discrete-variable quantum systems}

We now look into ways to measure the phase-space distribution of states that are hybridizations of continuous- and discrete-variable systems.
Many experiments and processes that consider the state of qubits also implicitly involve the presence of quantum light.
Likewise, many optical experiments also involve the interaction with some kind of atom, whether or not artificial.
This interaction is the foundation of quantum electrodynamics and circuit quantum electrodynamics.

Conversely, the continuous-variable Wigner function can serve as the position and momentum of an atom, so creating a composite function of discrete a continuous variables may be interpreted as the atomic state depending on its position and momentum.
This was shown useful to consider in \Refe{Davies2018} when looking at the full Wigner function of an atom.
By taking this composite approach there is much insight to be gained, from properties of light-matter interaction to other phenomena such as spin-orbit coupling.
In this section we will focus on the former application as it more directly applies to use in quantum technologies.

There are many ways phase-space methods can be used in discerning correlations that arise in the interaction between continuous- and discrete-variable systems.
Like in \Eq{MarginalWigner} for two qudits, it is usual to simply consider the marginal functions for each system individually, tracing out either the discrete-variable system or the continuous-variable system.
This is also implicitly the case in measuring the Wigner function in qubit experiments when using quantum states of light to manipulate qubits.
In theory, the qubit state should not have any correlations with the optical state, however there will always be some leakage, interpreted as decoherence in the qubit.

On the other side, it is typical to just consider the optical state in a light-matter interaction.
An important example is the generation of an optical Schr\"odinger cat state. 
In such experiments, it is normal to entangle light with a two-level atom, by doing this one can then split the light into the superposition of two coherent states, generating a Schr\"odinger cat state.\Cite{Deleglise:2008gt}.
At the point of measurement of the optical state, the goal is to produce separable atomic and optical states, and therefore minimizing decoherence in the Wigner function.
Like in the previous case, there may be some correlations that exist between the qubit and the field, therefore it may be more desirable to have an idea of the correlations that exist between the two systems.

\begin{figure}
	\includegraphics[width = \linewidth]{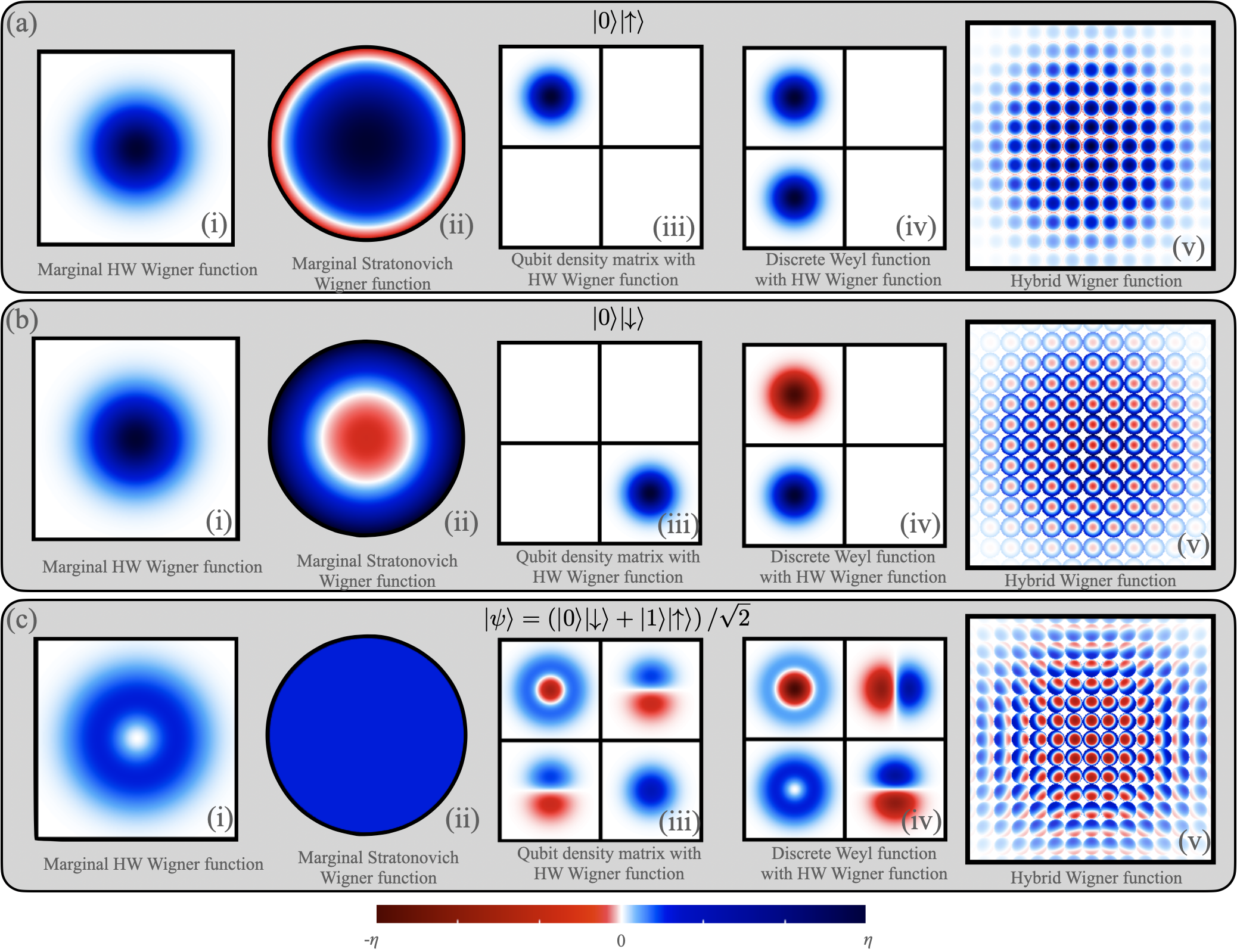}
	\caption{\label{HybridWignerFunctions}Examples of hybrid continuous and discrete variable systems in phase space.
	We show various ways to display different states in phase space.
	\Sub{a} shows different representations of $\ket{0}\ket{\uparrow}$.
	Likewise \Sub{b} shows different representations of $\ket{0}\ket{\uparrow}$
	The last state is a hybrid Bell state $(\ket{0}\ket{\downarrow}+\ket{1}\ket{\uparrow})/\sqrt{2}$ which is shown in \Sub{c}.
	For each state there are five different plots.
	In each \Sub{i} we show the marginal function for the HW Wigner function, with colorbar values $\eta=2$, similarly in \Sub{ii} is the marginal Stratonovich Wigner function for the qubit where the colorbar values are $\eta=(1+\sqrt{3})/2$.
	We can see that for the two separable states in \Sub{a} and \Sub{b} that the two marginal functions are pure states, conversely we see in \Sub{c} that the entangled state produces mixed states in the marginal functions, where the qubit has a uniform value over the distribution.
	Note that the qubit has been plotted in a Lambert azimuthal area preserving map, see text for details.
	In \Sub{iii} we have plotted the density matrix for the qubit, where in each entry we show the HW Wigner function.
	Similarly in \Sub{iv} we have a composition of the discrete Weyl function where in each entry the HW Wigner function has been plotted, note that this corresponds to taking Pauli measurements of the qubit and then plotting the HW Wigner function, as a result the marginal HW Wigner function is appears as the identity measurement of the qubit.
	In both \Sub{iii}  and \Sub{iv} the colorbar values are $\eta=2$.
	Finally in \Sub{v} we show a hybrid representation developed in \Refe{Davies2018,Rundle2019} which have colorbar values $\eta = 1 + \sqrt{3}$
	}
\end{figure}

Different examples of how one can plot such hybrid states are shown in \textbf{\Fig{HybridWignerFunctions}}.
We present some theoretical methods displayed in \Fig{HybridWignerFunctions}.
We will then discuss how some of these methods have been implemented experimentally and the future possibilities of measurement of hybrid quantum systems.

We will first give examples of the results when taking the marginal Wigner functions, calculated 
\begin{eqnarray}
	W(\alpha) &=&  \frac{1}{4\pi} \int_0^{2\pi}\int_0^{\pi} W(\alpha,\theta,\phi) \sin\theta\,\ud\theta\ud\phi\\
	W(\theta,\phi) &=&  \frac{1}{2\pi} \int_{-\infty}^{\infty} W(\alpha,\theta,\phi) \,\ud^2\alpha.
\end{eqnarray}
For the separable states, $\ket{0}\ket{\uparrow}$ and $\ket{0}\ket{\downarrow}$, shown in \FigSub{HybridWignerFunctions}{a} and~\Sub{b}, the marginal functions are as expected, in both cases the HW Wigner function is the vacuum state and the qubit is in the $\ket{\uparrow}$ and $\ket{\downarrow}$ state respectively.
\FigSub   {HybridWignerFunctions}{c} show the marginal Wigner functions for the state $(\ket{0}\ket{\downarrow}+\ket{1}\ket{\uparrow})/\sqrt{2}$ -- a hybrid analog of a Bell state.
As a result, the quantum correlations are non-local and the marginal Wigner functions are mixed states.
Where the qubit state is maximally mixed, which can be seen in \FigSub   {HybridWignerFunctions}{c}\Sub{ii} with a constant value over the whole distribution.

Note that we have chosen an alternative projection to display the qubit state.
We have used the Lambert azimuthal equal-area projection,\Cite{Lambert} in order to display the qubit completely.
It is similar to a stereographic projection with the north pole in the center, however the south pole is now projected to a unit circle, rather than to infinity.
The function is then distributed around the disk to preserve the area -- which is an important property when considering probability distributions.
This results in the equator of the sphere existing on the concentric circle with radius $1/\sqrt{2}$.
For a more in-depth explanation of this projection for use in phase-space methods see the text and appendices of \Refe{RundlePhD}.

The Lambert azimuthal equal-area projection was also taken in \Refe{EverittCatt} where the authors also considered marginal Wigner function for both subsystems.
The work in \Refe{EverittCatt} consided the Tavis-Cummings interaction \cite{TavisCummings} between an optical cavity and an atomic $j=5/2$ state initially in an atomic Schr\"odinger cat state.
By considering the Wigner functions of both systems, it is then insightful to visualize the transfer of the Schr\"odinger cat states between the two systems.
Demonstrating how one can overcome decoherence in quantum systems by coupling a large atomic state to an optical mode.
However, this lacks the consideration of the non-local correlations between the two systems, and any entangled state will result in Wigner functions similar to \FigSub{HybridWignerFunctions}{c}\Sub{i} and~\Sub{ii}.

To get a sense of these non-local correlations a hybridization approach is needed.
One approach has been to hybridize a density matrix for a qubit with the Wigner function for the field mode.
This results in a $2 \times 2$ grid, representing the entries of the qubit density matrix, where on each entry of the density matrix the corresponding Wigner function for the field mode is shown.
We can label this Wigner function $w(i,j)$, where $i,j \in \{0,1\}$, such that 
\begin{equation}
	w(i,j) = \Trace{ \left(\ket{i}\bra{j}\otimes \HWPi(\HWvar)\right) \rho },
\end{equation}
where $\rho$ is the state for the full system.
Such an approach reveals much more about the non-local correlations between the two systems, and can be found in works such as \Refe{Morin2014,Huang2019}.
Examples of this approach are shown in \Fig{HybridWignerFunctions}\Sub{iii}.
Note that the real value has been taken, which doesn't make a difference for the first two states. 
However, $(\ket{0}\ket{\downarrow}+\ket{1}\ket{\uparrow})/\sqrt{2}$ has an imaginary component on the off-diagonal entries that has been ignored.
Since we're concerned with eigenstates of $\Sz$, the local terms are shown in diagonal entries and the non-local correlations can be seen in the off-diagonal entries, which is clearly seen in \FigSub{HybridWignerFunctions}{c}\Sub{iii}.
A similar approach was taken in \Refe{Agudelo2017} where one can generate a nonclassicality quasiprobability matrix to explore

Alternatively one could take Pauli measurements of the qubit, and then calculate the HW Wigner function in that basis, the resulting distribution is shown in \Fig{HybridWignerFunctions}\Sub{d} and can be considered as a hybridization of the discrete Weyl function with the HW Wigner function, where
\begin{equation}
	\mathcal{W}(z,x,\HWvar)=\Trace{ \left(\DWD_2(z,x)\otimes \HWPi(\HWvar)\right) \rho }.
\end{equation}  
Similarly to the density matrix approach this produces a $2\times 2$ grid, where this time the elements correspond to Wigner function, yielding a joint quasiprobability distribution.
For each state, we take an identity measurement, this is equivalent to taking the marginal HW Wigner function, where the equivalence can be considered with \FigSub{HybridWignerFunctions}{c}\Sub{i}.

Such a method has gained much interested in practice, where these exact states were considered in \Refe{Eichler2012}, where the authors took measurements in the Pauli basis at different points in position-momentum phase space.
By building up measurement statistics each point in phase space was then represented by the expectation value of the given Pauli operator at this point in phase space.
Similar results that consider entanglement between a qubit and a Schr\"odinger cat states have also shown how this method is useful for characterizing entanglement.~\cite{Hacker2019, Vlastakis2015}

Alternatively we can consider a different method to hybridize Wigner functions for the two systems.
One could hybridize the Wootters Wigner function with the HW Wigner function in a similar way to coupling the density matrix, as was done in \Refe{Costanzo2015}.
However here we want to consider hybridizing the Stratonovich-Kernel Wigner function with the the HW Wigner function.\Cite{Davies2018,Rundle2019}
At each point $\dot\HWvar$ we can plot the Wigner function for the qubit, producing a lattice of Wigner functions where
\begin{equation}
	W(\dot\HWvar,\theta,\phi) = \Trace{\left(\HWPi(\dot\HWvar)\otimes\SUNPi{2}{1}(\theta,\phi)\right) \rho }.
\end{equation}
We can then add transparency to each of the spheres according to the maximum value of the Wigner function at that point $\dot\HWvar$, $\max_{\theta,\phi}W(\dot\HWvar,\theta,\phi)$, resulting in an envelope over the distribution.
Examples of this are given in \Fig{HybridWignerFunctions}\Sub{a} -~\Sub{c} \Sub{v}, where for the separable states we can see the envelope of the vacuum state.
At every discrete point in position and momentum phase space we then see the $\ket{\uparrow}$ state, in \Fig{HybridWignerFunctions}\Sub{a}\Sub{v}, and the $\ket{\downarrow}$ state in \Fig{HybridWignerFunctions}\Sub{b}\Sub{v}.
Like with the Wootters Wigner function, when there are non-local correlations between the two systems, we see a dependence between the degrees of freedom.
Where in \Fig{HybridWignerFunctions}\Sub{c}\Sub{v} we can see a radial dependence on the state of the qubit.

All these methods provide different snapshots into such coupled systems and in combination build up an understanding into the of these types of non-local correlations.
We also see certain overlap between different methods of plotting such systems, which further allows us to get a sense of the overall picture of coupled quantum systems in phase space.

\section{Conclusions and Outlook}

In this review we have explored the phase-space formulation of quantum mechanics, considering how different systems can be represented in phase space.
Our focus was providing the framework to represent discrete-variable systems in phase space, in particular the different ways single- and multi-qubit states can be considered.

We provided the general framework to generate an informationally complete phase-space distribution from any arbitrary operator, as well as how to transform between any of the phase-space representations and how the dynamics of quantum systems can be modelled in phase space.
This was followed by examples of discrete systems in phase space, with a focus on the construction on qubit and multi-qubit states.
We have provided a link between between the Stratonovich kernel and the Wootters kernel for a qubit state.
Given the resulting structure of a simplex inside a sphere, this provides a path into alternative constructions of discrete Wigner function for qudits and symmetric informationally complete states, where we can consider an $N^2$-dimensional simplex inside the \SU{N} $N^2-1$-dimensional surface.

From the construction on Wigner functions for discrete quantum states, we then explored how such distributions can be useful for applications to quantum technologies.
Exploring how well-known metrics in quantum information can be written in the language of phase space, as well as methods to use them in experimental settings.
There are still more metrics than can be translated into phase space, and perhaps even more that are peculiar to the phase-space representation, such as the negative volume of the Wigner function.

We then considered direct measurement of quantum systems in phase space, beginning with the phase-space reconstruction with single-qubit states before treating more complicated composite systems.
Where we first looked at multi-qubit states, where the dimension of the Hilbert space increases as $2^N$, whereas the degrees of freedom for a phase-space distribution only scale as $2N$ for $N$ qubits.
Therefore such methods can provide computational simplifications in larger systems -- especially if we are only interested in finding certain types of correlations within a larger state.

We then moved onto the current work of hybrid quantum systems.
Although the distributions considered here and simple single-qubit states these provide an outline into how come can adapt phase-space methods for composite systems with heterogeneous degrees of freedom. 
We note that there is also some exploration into how this can be extended to consider multi-qubit states coupled to a field mode via a Tavis-Cummings coupling in \Refe{RundlePhD}.
Conversely we can consider three continuous-variable distributions to each qubit, acting as the position and momentum in three dimensions to each atom, such an approach was considered in \Refe{Davies2018} when treating the states of atoms.

\medskip

\section*{Acknowledgements}
\medskip
The authors would like to thank Todd Tilma for fruitful discussions.
RPR would like to thank B\'alint Koczor for interesting and informative discussions and bringing to our attention some important work, RPR would also like to thank Steffen Glaser and Robert Zeier for valuable feedback.
RPR also acknowledges support from the EPSRC grant number EP/T001062/1 (EPSRC Hub in Quantum Computing and Simulation).

\medskip

%
\bibliographystyle{unsrt}
\bibliography{refs.bib}

%
%


\end{document}